%% file: bec_fb_capacity.tex
\documentclass[conference,onecolumn]{IEEEtran}

\usepackage[margin=0.75in]{geometry}

\IEEEoverridecommandlockouts                              

\usepackage{amsmath, amssymb}
\usepackage{wasysym}
\usepackage{bbold}
\usepackage{tikz}
\usetikzlibrary{shapes,arrows,fit}
\usetikzlibrary{plotmarks}
\usetikzlibrary{positioning}
\usetikzlibrary{decorations.markings}

\usepackage{pgfplots}
 \pgfplotsset{compat=newest} 
\usepackage[none]{hyphenat}
\usepackage[normalem]{ulem} 
\usepackage{color}

\let\labelindent\relax
\usepackage{enumitem}

\newcommand{\G}{\text{G}}
\newcommand{\B}{\text{B}}
\newcommand{\GG}{\text{GG}}
\newcommand{\GB}{\text{GB}}
\newcommand{\BG}{\text{BG}}
\newcommand{\BB}{\text{BB}}
 \newcommand{\mc}[1]{\mathcal{#1}}

\newcommand{\ve}[1]{\uline{\smash #1}}

\newcommand{\truth}{\mbox{\textbb{1}}}

\newcommand{\epsone}[1]{\ensuremath{\epsilon_1(#1)}}			
\newcommand{\epstwo}[1]{\ensuremath{\epsilon_2(#1)}} 			
\newcommand{\epsj}[1]{\ensuremath{\epsilon_j(#1)}}			
\newcommand{\epsonetwo}[1]{\ensuremath{\epsilon_{12}(#1)}}		
\newcommand{\epsonenottwo}[1]{\ensuremath{\epsilon_{1\bar{2}}(#1)}}	
\newcommand{\epsnotonetwo}[1]{\ensuremath{\epsilon_{\bar{1}2}(#1)}}	
\newcommand{\epsnotonenottwo}[1]{\ensuremath{\epsilon_{\bar{1}\bar{2}}(#1)}}	

\newcommand{\flowdiv}[3]{\ensuremath{f_{#1}^{#2}(#3)}}		

\newtheorem{prop}{Proposition}
\newtheorem{thm}{Theorem}
\newtheorem{lemma}{Lemma}

\newtheorem{example}{Example}
\newtheorem{remark}{Remark}

\tikzstyle{chan} = [draw, rectangle, rounded corners,minimum height=2em, minimum width=4em]
\tikzstyle{block} = [draw, rectangle, minimum height=4em, minimum width=2em]
\tikzstyle{block2} = [draw, rectangle, minimum height=1.5em, minimum width=2em]
\tikzset{->-/.style={decoration={markings,mark=at position #1 with {\arrow{>}}},postaction={decorate}}}

\author{\IEEEauthorblockN{ Michael Heindlmaier\IEEEauthorrefmark{1}, Navid Reyhanian\IEEEauthorrefmark{2}, Shirin Saeedi Bidokhti\IEEEauthorrefmark{1}}  \thanks{The authors are in alphabetical order and contributed equally to the work.}
\\
\IEEEauthorblockA{\IEEEauthorrefmark{1}Institute for Communications Engineering, Technische Universit\"at M\"unchen,
Munich, Germany\\
\IEEEauthorrefmark{2}University of Tehran, Tehran, Iran\\
Email: michael.heindlmaier@tum.de, n.reyhanian@ut.ac.ir,  shirin.saeedi@tum.de}}

\begin{document}


\title{\vspace{.25in}On Capacity Regions of Two-Receiver Broadcast Packet Erasure Channels with Feedback and Memory}

\maketitle

\begin{abstract}
The two-receiver broadcast packet erasure channel with feedback and memory is studied. Memory is modeled using a finite-state Markov chain representing a channel state.
Outer and inner bounds on the capacity region are derived when the channel state is strictly causally known at the transmitter. The bounds are both formulated in terms of feasibility problems and they are matching in all but one of the constraints. The results are extended to feedback with larger delay. Numerical results show that the bounds are close in many examples and the gains offered through feedback can be quite large.  
The presented outer bound meets the inner bound recently derived in \cite{Kuo_Wang2014} and hence describes the capacity region.
\end{abstract}

\input{intro.tex}

\input{model.tex}

\input{bounds.tex}

\input{schemes.tex}


\input{results.tex}

\section{Conclusion}
We investigated the two-user broadcast packet erasure channel with feedback and memory. We modeled the channel memory by a finite state machine and found outer and inner bounds on the capacity region when the channel state is known strictly causally at the encoder. 
To achieve the inner bound we proposed a probabilistic scheme and presented a deterministic queue-length based algorithm. The results are extended to feedback with larger delay.
Numerical results show that the gains offered through feedback can be quite large and that the difference between the outer and inner bound is small.
One possible future direction is to determine for which cases the inner and outer bounds meet.

 \section*{Acknowledgments}
 The authors are supported by the German Ministry of Education and Research in the framework of the Alexander von Humboldt-Professorship and by the grant DLR@Uni of the Helmholtz Allianz.
The work of S. Saeedi Bidokhti was partially supported by the Swiss National Science Foundation Fellowship no. 146617.
 The authors would like to thank Gianluigi Liva and Andrea Munari for motivating the problem, and Gerhard Kramer for his helpful comments.
The authors are also grateful to Chih-Chun Wang for helpful discussions that helped identify a mistake in an earlier version of this work.

\newpage
\appendices

\input{appendix.tex}

\bibliographystyle{IEEEtran}
\bibliography{bib}

\end{document}

%% file: intro.tex
\section{Introduction}

The capacity of broadcast channels (BCs) remains unresolved both without and with feedback.
It was shown in \cite{gamal1978feedback} that feedback does not increase the capacity  of physically degraded BCs. 
Nevertheless, feedback increases the capacity of general BCs and even partial feedback can help \cite{dueck1980partial, kramer2003capacity}. Feedback also increases the capacity region of AWGN BCs \cite{ozarow1984broadcast,4655434}.

The capacity region of  memoryless broadcast packet erasure channels (BPECs) with feedback (FB) was found in \cite{georgiadis2009broadcast} for two receivers. The region is characterized by the closure of all non-negative rate pairs $(R_1,R_2)$ such that
\begin{align}
 &\frac{R_1}{1-\epsilon_1} + \frac{R_2}{1-\epsilon_{12} } \leq 1 \nonumber \\
& \frac{R_1}{1-\epsilon_{12}} + \frac{R_2}{1-\epsilon_2} \leq 1, \nonumber 
\end{align}
where $\epsilon_1$ and $\epsilon_2$ are the erasure probabilities at receiver $1$ and $2$, respectively, and $\epsilon_{12}$ is the probability of erasure at both receivers.
In particular, feedback increases the capacity and this is of practical interest  since the required feedback is only a low-cost ACK/NACK signal that is easy to implement in BPECs.

This result has been extended to certain cases of broadcast channels with more number of receivers in \cite{6522177,gatzianas2012feedback,wang2012capacity}. In all these works, the capacity region is achieved using feedback-based coding algorithms that are based on network coding ideas. The converse theorems are proved by proving genie-aided outer bounds on the capacity region. The trick is that the genie helps the receivers such that the broadcast channel becomes a physically degraded one, for which the capacity region with feedback is known \cite{gamal1978feedback, bergmans1973random, gallager1974capacity}.

The capacity region of two-receiver multiple-input BPECs with feedback has been studied in \cite{6847153} where the capacity region is derived and is shown to be achievable using linear network codes (LNC). The schemes are also applied to partially Markovian and partially controllable broadcast PECs where the linear network coding rate region is characterized by a linear program which exhaustively searches for the LNC scheme(s) with the best possible throughput.

In a recent trend of research, noisy feedback has been studied and achievable schemes are developed in \cite{shayevitz2013capacity, venkataramanan2010achievable}.

This paper studies BPECs with memory and feedback. The problem is motivated by the bursty nature of erasures in practical communication systems, e.g., satellite links \cite{lutz1991land, fontan2001statistical,ibnkahla2004high}. We model the memory of a channel by a finite state machine and a set of state-dependent erasure probabilities. For finite state channel models see e.g. \cite{sadeghi2008finite} and the references therein. 

When there is no feedback, one can use erasure correcting codes for memoryless channels in combination with interleavers to decorrelate the erasures. But feedback enables more sophisticated coding methods and
several such schemes are discussed in \cite{heindlmaier2014netcod}. We remark that \cite{dabora2010capacity} studied the general broadcast channel with feedback and memory and considered different cooperation scenarios. The capacity characterizations in \cite{dabora2010capacity} are, however, in multi-letter form and not computable.

The main contribution of this paper is to provide lower and upper bounds on the capacity region for two receivers when the channel state is strictly causally known at the transmitter. Both bounds are formulated in terms of feasibility problems, and are similar in all but one set of constraints.
Our outer bound is a genie-aided bound. The bound is subtle in the sense that it cannot be derived directly using the results of \cite{gamal1978feedback, bergmans1973random, gallager1974capacity}. 
Our proposed achievable scheme extends the queue-based algorithms of \cite{georgiadis2009broadcast,gatzianas2012feedback, neely2010stochastic} to incorporate knowledge about the past channel states. The techniques generalize to BPECs with delayed feedback. 

During the preparation of this work, we were informed that in a parallel line of work \cite{6847153, Kuo_Wang2014} investigated dynamic scheduling algorithms for a similar problem. The outer bound we derive in Section~\ref{sec:UB} matches the inner-bound derived in \cite{Kuo_Wang2014} and thus characterizes the capacity region.
The inner bound we derive in Section \ref{sec:ach} is included in this region, and this inclusion can be strict.

This paper is organized as follows:
We introduce notation and the system model in Section \ref{sec:model}, and elaborate on the main result in Section~\ref{sec:main}. The outer bound  is presented in Section \ref{sec:UB} and the inner bound and two achievable schemes are discussed in Section \ref{sec:ach}. In Section \ref{sec:comparison} we discuss implications of our results.

%% file: model.tex
\section{Notation and System Model}
\label{sec:model}

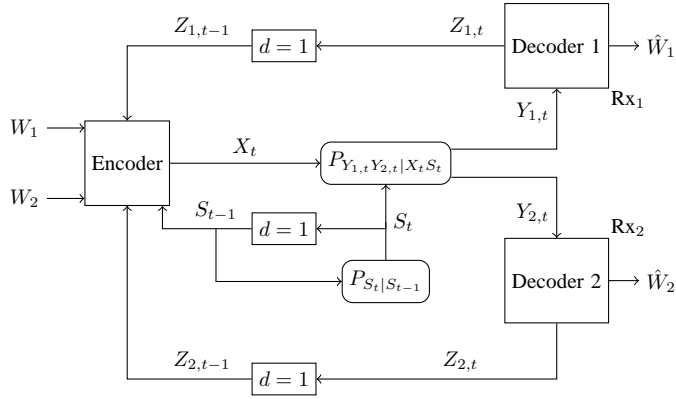
\begin{figure}[t]
\centering
\begin{tikzpicture}[node distance=5mm, scale=1, every node/.style={scale=0.8}]
\input{./figs/block_diag.tex}
\end{tikzpicture}
\caption{Block diagram for the broadcast packet erasure channel with visible state. The box marked with $d=1$ represents a delay of one time unit.}
\label{fig:block}
\end{figure}

\subsection{Notation}
Random variables are denoted by capital letters.
A finite sequence (or string) of random variables $X_1,X_2,\ldots,X_n$ is denoted by $X^n$. In this context, sequences always refer to sequences in time.
Sequences may have subscripts, e.g. $X_j^n$ denotes $X_{j,1},X_{j,2},\ldots,X_{j,n}$.
It is sometimes convenient to collect random variables that appear at the same time in a vector.
Vectors are written with underlined letters, e.g.,  $\ve Z_t=(Z_{1,t}, Z_{2,t})$.
Sets are denoted by calligraphic letters, e.g., $\mc X$.
The indicator function $\truth\{\cdot \}$ takes on the value $1$ if the event inside the brackets is true and $0$ otherwise. 
The probability of a random variable $X$ taking on a realization $x$ given an event $\mc E$ is written as $\Pr[X=x|\mc E]$. Often, the conditional event corresponds to another random variable $Y$ taking on some realization $y$. This conditional probability is written as $\Pr[X=x|Y=y]$ or equivalently $P_{X|Y}(x|y)$. 
The equivalent expressions $\Pr[X|Y]$ or $P_{X|Y}$ are used to address the conditional probability (distribution) for any outcome of $X,Y$.

The conditional expectation of a function $f$ of a random variable $X$ given another random variable $Z$ is itself a random variable and is written as $\mathbb E[f(X)|Z]$. Using the law of total expectation $\mathbb E[f(X)|Z] = \mathbb E\bigl[\mathbb E[f(X)|YZ] \big| Z \bigr]$. Note that if $X-Y-Z$ forms a Markov chain, we can write $\mathbb E[f(X)|Z] = \mathbb E\bigl[\mathbb E[f(X)|Y] \big| Z \bigr]$.

\subsection{System Model}
A transmitter wishes to communicate two independent messages $W_1$ and $W_2$ (of $nR_1$, $nR_2$ packets, respectively) to two receivers $\text{Rx}_1$ and $\text{Rx}_2$ over $n$ channel uses. Communication takes place over a packet erasure broadcast channel with memory and feedback as described below:

The input to the broadcast channel  at time $t$, $t=1,\ldots,n$, is denoted by $X_t \in \mc X$.
The channel inputs correspond to packets of $L$ bits; we may represent this by choosing $\mc X = \mathbb{F}_{q}$ with $q=2^L$, and $L \gg 1$.
Transmission rates are measured in packets per slot and so entropies and mutual information terms are considered with logarithms to the base $q$. 

The channel outputs at time $t$ are written as $Y_{1,t}\in\mathcal{Y}$ and $Y_{2,t}\in\mathcal{Y}$ where $\mathcal{Y}=\mc X \cup \{E\}$.  Each $Y_{j,t}$, $j\in\{1,2\}$, is either $X_t$ (i.e., received perfectly) or $E$ (i.e., erased).

We define binary random variables $Z_{j,t}$, $j\in\{1,2\}$, $t=1,\ldots,n$, to indicate if an erasure occurred at receiver $j$ in time $t$; i.e., $Z_{j,t}=\truth\{Y_{j,t}=E\}$.
Clearly, $Y_{j,t}$ can be expressed as a function of $X_{t}$ and $Z_{j,t}$. Furthermore, $Y_{j,t}$ also determines $Z_{j,t}$. We denote $(Z_{1,t},Z_{2,t})$ by $\ve Z_t$.

The broadcast channel we study has memory that is modeled via a finite state machine with state $S_t$ at time $t$. The state evolves according to an irreducible aperiodic finite state Markov chain with state space $\mc S$ and steady-state distribution $\pi_s$, $s \in \mc S$. The initial state $S_0$ is distributed according to $\pi$.
Depending on the current random state of the channel, the channel erasure probabilities are specified through
the conditional distribution $P_{\ve Z_{t}|S_{t}}$. 
Arbitrary correlation between $(Z_{1,t},Z_{2,t})$ is permitted.
The transition probabilities between channel states are known at the transmitter.
Note that the sequence $\ve Z^n$ is correlated in time in general, hence the channel has memory.

After each transmission, an ACK/NACK feedback is available at the encoder from both receivers. Two possible setups can be considered for the encoding function $f_t$: 
\begin{itemize}
 \item[(i)] Only ACK/NACK feedback is available at the encoder:
\begin{align}
X_t = f_t(W_1,W_2,Z_1^{t-1},Z_2^{t-1}).
\end{align}
 \item[(ii)] ACK/NACK and the previous state feedback is known:
\begin{align}
X_t = f_t(W_1,W_2,Z_1^{t-1},Z_2^{t-1},S^{t-1}).
\end{align}
\end{itemize}

\begin{figure}
\centering
  \tikzset{>=latex}
\begin{tikzpicture}[scale = 0.5, every node/.style={scale=0.5}]
\input{./figs/fdg.tex}
 \end{tikzpicture}
\caption{Bayesian network for the two-receiver broadcast packet erasure channel with memory and ACK/NACK + previous state feedback (visible state), for $n=3$ and $d=1$. Dependencies due to feedback are drawn with dashed lines.}
\label{fig:fdg_system}
\end{figure}
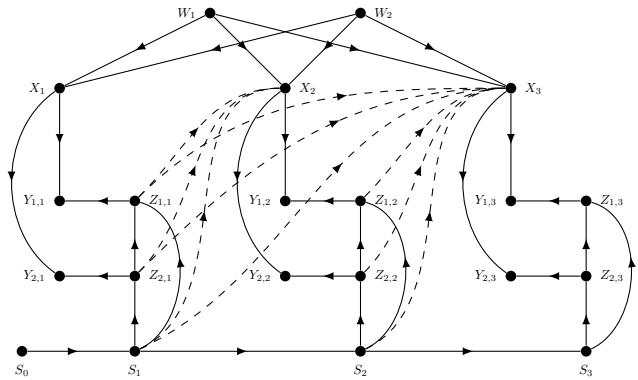

Depending on whether the transmitter knows the previous channel state or not, we call the state \emph{visible} or \emph{hidden}. This paper is focused on the problem with visible states (see Fig.~\ref{fig:block}).
The joint probability mass function of the system then factorizes as
\begin{align}
&P_{W_1 W_2 X^n S^n Y_1^n Y_2^n Z_1^n Z_2^n} = 
P_{W_1} P_{W_2} P_{S_0} \prod_{t=1}^n P_{S_t|S_{t-1}} P_{X_t|S^{t-1} Z_1^{t-1} Z_2^{t-1}} P_{Z_{2,t}|S_t} P_{Z_{1,t}|Z_{2,t}S_t} P_{Y_{1,t}|X_t Z_{1,t}} P_{Y_{2,t}|X_t Z_{2,t}}. \nonumber
\end{align}
The corresponding Bayesian network\footnote{The Bayesian network can also easily be transformed into a functional dependency graph (FDG) \cite{kramer2003capacity}, for which simple rules for checking conditional independence exist.} for the visible case is shown in Fig.~\ref{fig:fdg_system} and can be used to determine conditional independence of random variables.

The state can be visible either because it is explicitly available at the transmitter or because it may be determined from the available feedback. The latter is illustrated via the following example. 

\begin{example}
\label{exampleGilbert}
Consider a Gilbert-Elliot model \cite{gilbert1960capacity,elliott1963estimates} with state space $\mc S = \{\GG,\GB, \BG, \BB\}$ where $\G$ and $\B$ respectively refer to a good and bad state at each user. 

One case of interest is when we have erasure in state $\B$ and no erasure in state $\G$, i.e.,
\begin{align}
&P_{\ve Z_t|S_t}(0,0|\GG)=1,\quad P_{\ve Z_t|S_t}(0,1|\GB)=1\nonumber\\
&P_{\ve Z_t|S_t}(1,0|\BG)=1,\quad P_{\ve Z_t|S_t}(1,1|\BB)=1. \label{eq:param_gilbert}
\end{align}
In such a channel, the feedback $\ve Z_t$ determines the channel state, and we thus say that the state is visible. We use this channel model for our simulation results in Section~\ref{sec:comparison}.
\end{example}

We define the probability of erasure events given the \emph{previous} channel state $s$ as follows:
\begin{align}
 \epsonetwo{s}&=P_{\ve Z_t|S_{t-1}}(1,1|s),\quad
 &\epsonenottwo{s}&=P_{\ve Z_t|S_{t-1}}(1,0|s),\nonumber\\
 \epsilon_{\bar 12}(s)&=P_{\ve Z_t|S_{t-1}}(0,1|s),\quad  
  &\epsnotonenottwo{s} &=P_{\ve Z_t|S_{t-1}}(0,0|s),\nonumber\\
   \epsone{s}&=\epsonetwo{s}+\epsonenottwo{s},\quad 
  &\epstwo{s}&=\epsonetwo{s}+\epsnotonetwo{s}. \label{eq:def_epsilon}
\end{align}
Note that these probabilities do not depend on $t$ in our setup.

The goal is to have each decoder $\text{Rx}_j$ reliably estimate $\hat W_j = h_j(Y_j^n)$ from its received sequence $Y_j^n$.
A rate-pair $(R_1,R_2)$ is said to be achievable if the error probability $\Pr[\hat W_1 \neq W_1, \hat{W}_2\neq W_2]$ can be made arbitrarily small as $n$ gets large.
The capacity region $\mc C_{\text{ fb}}^\text{mem}$ is the convex closure of the achievable rate pairs.

\section{Main Result}
\label{sec:main}
The main result of this paper is the following bounds on the capacity region of the two-user packet erasure broadcast channel with memory and ACK/NACK feedback.

Define $\bar{\mc C}_{\text{ fb}}^\text{mem}$ to be  the closure of rate pairs $(R_1,R_2)$ for which there exist variables $x_s$, $y_s$, $s\in \mc S$ such that
 \begin{align}
 &0\leq x_s\leq 1,\quad 0\leq y_s \leq 1 \label{eq:posouter}\\
  &R_1  \leq \sum_{s \in \mc S} \pi_s (1-\epsone{s}) x_s  \label{eq:R1_constr1outer}\\
  &R_1  \leq \sum_{s \in \mc S} \pi_s (1- \epsonetwo{s}) (1-y_s) \label{eq:R1_constr2outer} \\
  &R_2  \leq \sum_{s \in \mc S} \pi_s (1-\epstwo{s}) y_s  \label{eq:R2_constr1outer} \\
  &R_2  \leq \sum_{s \in \mc S} \pi_s (1-\epsonetwo{s}) (1-x_s).  \label{eq:R2_constr2outer}
 \end{align}

Define, furthermore, $\underline{\mc C}_{\text{ fb}}^\text{mem}$ to be  the closure of rate pairs $(R_1,R_2)$ for which there exist variables $x_s$, $y_s$, $s\in \mc S$ such that
 \begin{align}
 &0\leq x_s\leq 1,\quad 0\leq y_s \leq 1 \label{eq:pos}\\
 &x_s + y_s \geq 1,~\forall s \in \mc S \label{eq:pa3_pos}\\
  &R_1  \leq \sum_{s \in \mc S} \pi_s (1-\epsone{s}) x_s  \label{eq:R1_constr1}\\
  &R_1  \leq \sum_{s \in \mc S} \pi_s (1- \epsonetwo{s}) (1-y_s) \label{eq:R1_constr2} \\
  &R_2  \leq \sum_{s \in \mc S} \pi_s (1-\epstwo{s}) y_s  \label{eq:R2_constr1} \\
  &R_2  \leq \sum_{s \in \mc S} \pi_s (1-\epsonetwo{s}) (1-x_s).  \label{eq:R2_constr2}
 \end{align}

Note that $\bar{\mc C}_{\text{ fb}}^\text{mem}$ and $\underline{\mc C}_{\text{ fb}}^\text{mem}$ differ in \eqref{eq:pa3_pos}.
\begin{thm}
\label{thm:capacity}
 The capacity region $\mc C_{\text{ fb}}^\text{mem}$ of the two-user broadcast packet erasure channel with feedback and visible state is sandwiched between  $\underline{\mc C}_{\text{ fb}}^\text{mem}$ and $\bar{\mc C}_{\text{ fb}}^\text{mem}$; i.e,
 \begin{align}
  \underline{\mc C}_{\text{ fb}}^\text{mem} \subseteq \mc C_{\text{ fb}}^\text{mem}\subseteq\bar{\mc C}_{\text{ fb}}^\text{mem}.
 \end{align}
\end{thm}
For example, consider Theorem~\ref{thm:capacity} when $\mathcal{S}$ has one state only, say state $s$, which models a memoryless erasure broadcast channel. One may verify that the two regions $\underline{\mc C}_{\text{ fb}}^\text{mem}$ and $\bar{\mc C}_{\text{ fb}}^\text{mem}$ match, and that eliminating variables $x_s,y_s$, the well-known result of \cite{georgiadis2009broadcast} follows. Let us call this capacity region $\mc C_\text{fb}(s)$. Now consider the case where $|\mc S|$ is larger: One might guess that the capacity region ${\mc C}_{\text{ fb}}^\text{mem}$ is the average direct sum (set sum) of the capacity regions $\mc C_\text{fb}(s)$ over all states $s\in\mathcal{S}$. However, this is \textit{not} in general the case: the capacity region can be strictly larger than the average direct sum of the $\mc C_\text{fb}(s)$. We expand on this remark in Section~\ref{sec:comparison}.

In Section~\ref{sec:UB}, we prove that $\bar{\mc C}_{\text{ fb}}^\text{mem}$ forms an outer bound; i.e., for any achievable scheme the problem defined in \eqref{eq:posouter} - \eqref{eq:R2_constr2outer} is feasible. This is done by bounding the achievable rates $R_1,R_2$ and expressing them in a manner similar to \eqref{eq:posouter} - \eqref{eq:R2_constr2outer}. Our converse proof is motivated by \cite{bergmans1973random,gamal1978feedback,DanaHassibi05}.

In Section \ref{sec:ach}, we introduce two schemes that can achieve any rate-pair in $\underline{\mc C}_{\text{ fb}}^\text{mem}$, and thus prove achievability of it. The first scheme is a probabilistic scheme that chooses encoding operations according to a probability distribution.
The second scheme uses a deterministic queue-length based algorithm that chooses encoding operations based on the feedback and the current buffer states.
This scheme stabilizes all queues in the network for every rate pair in $\underline{\mc C}_{\text{ fb}}^\text{mem}$.

While $\underline{\mc C}_{\text{ fb}}^\text{mem}$ and $  \bar{\mc C}_{\text{ fb}}^\text{mem}$ match and characterize the capacity in several examples, there are interesting cases where $  \underline{\mc C}_{\text{ fb}}^\text{mem}$ is strictly smaller than $  \underline{\mc C}_{\text{ fb}}^\text{mem}$. One such example is given in \cite[Sec. II.B]{Kuo_Wang2014}. In this example $\underline{\mc C}_{\text{ fb}}^\text{mem}$ turns out to be strictly smaller than the capacity region.

In Section \ref{sec:comparison}, we plot our inner and outer bounds on $\mc C_{\text{ fb}}^\text{mem}$ for a few examples and address the gain due to feedback and causal knowledge of the channel state. We furthermore discuss the gap between the inner and outer bounds and show that the capacity region is strictly larger that the average direct sum of the $\mc C_\text{fb}(s)$. Finally, we discuss variations of the problem with delayed feedback.

%% file: figs/block_diag.tex

\node at (1.5,0) [block] (Enc) {Encoder};
\node [chan, right = 20mm of Enc.east] (channel) {$P_{Y_{1,t} Y_{2,t}|X_t S_t}$};

\node [block, above right = 10mm of channel.north east] (Dec1) {Decoder 1};
\node [below right=-1mm of Dec1.south east] (Rx1) {Rx$_1$};
\node [right = 4mm of Dec1.east] (W1hat) {$\hat W_1$};

\node [block, below right = 10mm of channel.south east] (Dec2) {Decoder 2};
\node [above right=-1mm of Dec2.north east] (Rx2) {Rx$_2$};
\node [right = 4mm of Dec2.east] (W2hat) {$\hat W_2$};

\node [block2, left = 25mm of Dec1.west] (D1) {$d=1$};
\node [block2, below= 40mm of D1] (D2) {$d=1$};

\node [block2, below= 20mm of D1] (D3) {$d=1$};

\node at (Enc.north west) (Anch1) {};
\node at (Enc.south west) (Anch2) {};
\node at (channel.north east) (Anch3) {};
\node at (channel.south east) (Anch4) {};

\node (W1) [left= 5mm of Anch1.south] {$W_1$};
\node (W2) [left= 5mm of Anch2.north] {$W_2$};

\node at (Enc.south east) (Anch5) {};

\node [chan, below = 10mm of channel] (state_chan) {$P_{S_t|S_{t-1}}$};

\draw[->] (W1) -- (Anch1.south);
\draw[->] (W2) -- (Anch2.north);
\draw[->] (Enc) -- node[above](X){$X_t$} (channel);
\draw[->] (Anch3.south) -| node[left,pos=0.8] (Y1) {$Y_{1,t}$} (Dec1);
\draw[->] (Anch4.north) -| node[left,pos=0.8] (Y2) {$Y_{2,t}$} (Dec2);
\draw[->] (Dec1.west) -- node[above,pos =0.2] (Z1) {$Z_{1,t}$} (D1);
\draw[->] (Dec2.south) |- node[above,pos =0.7] (Z2) {$Z_{2,t}$}  (D2);
\draw[->] (D1) -| node[above,pos =0.2] (Z1del) {$Z_{1,t-1}$} (Enc);
\draw[->] (D2) -| node[above,pos =0.2] (Z2del) {$Z_{2,t-1}$} (Enc);
\draw[->] (Dec1) -- (W1hat);
\draw[->] (Dec2) -- (W2hat);

\draw[->] (state_chan) -- node[right] (S) {$S_{t}$} (channel.south);
\draw[->] (S.west) |- (D3);

\draw[->] (D3) -| node[above,pos =0.2] (Sdel) {$S_{t-1}$} (Anch5.west);
\draw[->] (Sdel) |- (state_chan);

%% file: figs/fdg.tex

\node at (4,0) [fill, circle, inner sep = 1mm] (W1) {};
\node [left=0.5mm of W1] () {$W_1$};

\node at (8,0) [fill, circle, inner sep = 1mm] (W2) {};
\node [right=0.5mm of W2] () {$W_2$};

\node at (0,-2) [fill,circle, inner sep = 1mm] (X1) {};
\node [left=0.5mm of X1] () {$X_1$};

\node at (6,-2) [fill,circle, inner sep = 1mm] (X2) {};
\node [right=0.5mm of X2] () {$X_2$};

\node at (12,-2) [fill,circle, inner sep = 1mm] (X3) {};
\node [right=0.5mm of X3] () {$X_3$};

\node at (0,-5) [fill,circle, inner sep = 1mm] (Y11) {};
\node [left=0.5mm of Y11] () {$Y_{1,1}$};

\node at (6,-5) [fill,circle, inner sep = 1mm] (Y12) {};
\node [left=0.5mm of Y12] () {$Y_{1,2}$};

\node at (12,-5) [fill,circle, inner sep = 1mm] (Y13) {};
\node [left=0.5mm of Y13] () {$Y_{1,3}$};

\node at (2,-5) [fill,circle, inner sep = 1mm] (Z11) {};
\node [right=0.5mm of Z11] () {$Z_{1,1}$};

\node at (8,-5) [fill,circle, inner sep = 1mm] (Z12) {};
\node [right=0.5mm of Z12] () {$Z_{1,2}$};

\node at (14,-5) [fill,circle, inner sep = 1mm] (Z13) {};
\node [right=0.5mm of Z13] () {$Z_{1,3}$};

\node at (0,-7) [fill,circle, inner sep = 1mm] (Y21) {};
\node [left=0.5mm of Y21] () {$Y_{2,1}$};

\node at (6,-7) [fill,circle, inner sep = 1mm] (Y22) {};
\node [left=0.5mm of Y22] () {$Y_{2,2}$};

\node at (12,-7) [fill,circle, inner sep = 1mm] (Y23) {};
\node [left=0.5mm of Y23] () {$Y_{2,3}$};

\node at (2,-7) [fill,circle, inner sep = 1mm] (Z21) {};
\node [right=0.5mm of Z21] () {$Z_{2,1}$};

\node at (8,-7) [fill,circle, inner sep = 1mm] (Z22) {};
\node [right=0.5mm of Z22] () {$Z_{2,2}$};

\node at (14,-7) [fill,circle, inner sep = 1mm] (Z23) {};
\node [right=0.5mm of Z23] () {$Z_{2,3}$};

\node at (-1,-9) [fill,circle, inner sep = 1mm] (S0) {};
\node [below=0.5mm of S0] () {$S_{0}$};

\node at (2,-9) [fill,circle, inner sep = 1mm] (S1) {};
\node [below=0.5mm of S1] () {$S_{1}$};

\node at (8,-9) [fill,circle, inner sep = 1mm] (S2) {};
\node [below=0.5mm of S2] () {$S_{2}$};

\node at (14,-9) [fill,circle, inner sep = 1mm] (S3) {};
\node [below=0.5mm of S3] () {$S_{3}$};

\draw[->-=.5] (W1) -- (X1);
\draw[->-=.5] (W1) -- (X2);
\draw[->-=.5] (W1) -- (X3);

\draw[->-=.5] (W2) -- (X1);
\draw[->-=.5] (W2) -- (X2);
\draw[->-=.5] (W2) -- (X3);

\draw[->-=.5] (X1) -- (Y11);
\draw[->-=.5] (X2) -- (Y12);
\draw[->-=.5] (X3) -- (Y13);

\draw[->-=.5] (X1) to [out=215,in=145] (Y21);
\draw[->-=.5] (X2) to [out=215,in=145] (Y22);
\draw[->-=.5] (X3) to [out=215,in=145] (Y23);

\draw[->-=.5] (Z11) -- (Y11);
\draw[->-=.5] (Z12) -- (Y12);
\draw[->-=.5] (Z13) -- (Y13);

\draw[->-=.5] (Z21) -- (Y21);
\draw[->-=.5] (Z22) -- (Y22);
\draw[->-=.5] (Z23) -- (Y23);

\draw[dashed,->-=.5] (Z11)  to [out=45,in=180] (X2);
\draw[dashed,->-=.6] (Z11) to [out=45,in=180] (X3);

\draw[dashed,->-=.5] (Z21) to [out=45,in=180] (X2);
\draw[dashed,->-=.6] (Z21) to [out=45,in=180] (X3);

\draw[dashed,->-=.5] (Z12) to [out=45,in=180] (X3);
\draw[dashed,->-=.5] (Z22) to [out=45,in=180] (X3);

\draw[dashed,->-=.5] (S1) to [out=25,in=180] (X2);
\draw[dashed,->-=.6] (S1) to [out=25,in=180] (X3);

\draw[dashed,->-=.5] (S2) to [out=25,in=180] (X3);

\draw[->-=.5] (S1) -- (Z21);
\draw[->-=.5] (S2) -- (Z22);
\draw[->-=.5] (S3) -- (Z23);

\draw[->-=.6] (S1) to [out=25,in=345] (Z11);
\draw[->-=.6] (S2) to [out=25,in=345] (Z12);
\draw[->-=.6] (S3) to [out=25,in=345] (Z13);

\draw[->-=.5] (Z21) -- (Z11);
\draw[->-=.5] (Z22) -- (Z12);
\draw[->-=.5] (Z23) -- (Z13);

\draw[->-=.5] (S0) -- (S1);
\draw[->-=.5] (S1) -- (S2);
\draw[->-=.5] (S2) -- (S3);

%% file: bounds.tex
\section{The Converse}
\label{sec:UB}
In this section, we prove that $\bar{\mc {C}}_{\text{ fb}}^\text{mem}$ is an outer bound on the capacity region. The general idea is to show that for any achievable scheme, there are parameters $x_s,y_s$, $s\in\mathcal{S}$, as in \eqref{eq:posouter} - \eqref{eq:R2_constr2outer}. We find these parameters by relating them to mutual information terms.

In order to bound $R_1$ and $R_2$, for any $\delta>0$, we  write the following  multi-letter bounds and  single-letterize them properly next. 
For $j\in\{1,2\}$, we define $\bar{j} \in \{1,2\}$ such that $\bar{j}\neq j$.
\begin{align}
nR_j&\leq I(W_j;Y_j^n)+n\delta \label{multi1}\\
nR_j&\leq I(W_j;Y_1^nY_2^n|W_{\bar{j}})+n\delta\label{multi3}
\end{align}
In \eqref{multi1} - \eqref{multi3}, we have used the independence of the messages and Fano's inequality \cite[Chapter 2.10]{cover2006elements}.

For $j=1$, the single-letterization is done as follows:
\allowdisplaybreaks
\begin{align}
R_1-\delta&\leq \frac{1}{n}I(W_1;Y_1^n)\nonumber \\
&\leq \frac{1}{n}I(W_1;Y_1^nS^{n-1})\nonumber \\
&=\frac{1}{n}\sum_{t=1}^n I(W_1;Y_{1,t}S_{t-1}|Y_1^{t-1}S^{t-2})\nonumber \\
&=\frac{1}{n}\sum_{t=1}^n \left[I(W_1;S_{t-1}|Y_1^{t-1}Z_1^{t-1}S^{t-2}) + I(W_1;Y_{1,t}|Y_1^{t-1}S^{t-1})\right]\nonumber \\
&\stackrel{(a)}{=}\sum_{t=1}^n \frac{1}{n}I(W_1;Y_{1,t}|Y_1^{t-1}S^{t-1})\nonumber \\
&{\leq}\sum_{t=1}^n \frac{1}{n}I(W_1Y_1^{t-1}S^{t-1};Y_{1,t}|S_{t-1})\nonumber \\
&\stackrel{(b)}{=}\sum_{t=1}^n \frac{1}{n}I(U_{1,t};Y_{1,t}|S_{t-1})\nonumber \\
&\stackrel{(c)}{=} I(U_{1,T};Y_{1,T}|S_{T-1}T)\nonumber \\
&= \sum_{s\in\mathcal{S}}\pi_s I(U_{1,T};Y_{1,T}|T,S_{T-1}=s).\label{no1}
\end{align}
In the above chain of inequalities, $(a)$ follows because $Z_1^{t-1}$ is a function of $Y_1^{t-1}$ and because of the Markov chain
\begin{displaymath}
 W_1-Y_1^{t-1}Z_1^{t-1}S^{t-2}-S_{t-1},
\end{displaymath}
$(b)$ follows by defining $U_{1,t}=(W_1Y_{1}^{t-1}S^{t-1})$, and  
\\$(c)$ follows by a standard random time sharing argument with time sharing random variable $T$.

Similarly, one obtains 
\begin{align}
R_1-\delta&\leq \frac{1}{n}I(W_1;Y_1^nY_2^n|W_2)\nonumber\\
&\leq\sum_{s\in\mathcal{S}}\pi_s I(U_{1,T};Y_{1,T}Y_{2,T}|U_{2,T}V_{T}T,S_{T-1}=s),\label{no2}
\end{align}
where $U_{2,T}=(W_2Y_{2}^{T-1}S^{T-1})$ and $V_{T}=(Y_1^{T-1}Y_{2}^{T-1}S^{T-1})$.

By symmetry, we also have the following bounds:
\begin{align}
R_2-\delta&\leq \sum_{s\in\mathcal{S}}\pi_s I(U_{2,T};Y_{2,T}|T,S_{T-1}=s)\label{no4}\\
R_2-\delta&\leq \sum_{s\in\mathcal{S}}\pi_s I(U_{2,T};Y_{1,T}Y_{2,T}|U_{1,T}V_TT,S_{T-1}=s). \label{no5}
\end{align}

\begin{remark} 
\label{remark1}
Note that 
\begin{itemize}
\item[(i)] $V_T$ is a  function of $(U_{1,T}U_{2,T})$, and
\item[(ii)] $\underline{Z}_T-TS_{T-1}-U_{1,T}U_{2,T}V_TX_T$ forms a Markov chain.
\end{itemize}
\end{remark}

The following lemma extends \cite[Lemma 1]{DanaHassibi05} and is proven in 
Appendix \ref{appendix:lemma1}.
\begin{lemma}
\label{lemma1}
For every $s\in\mathcal{S}$ and $j\in\{1,2\}$, we have:
\begin{alignat}{2}
I(U_{j,T};Y_{j,T}|T,S_{T-1}=s)&= (1-\epsj{s}) I(U_{j,T};X_{T}|T,S_{T-1}=s)\label{eq:lem1eq1}\\
I(U_{j,T};Y_{1,T}Y_{2,T}|U_{\bar{j},T} V_T T,S_{T-1}=s)&= (1-\epsonetwo{s}) I(U_{j,T};X_T|U_{\bar{j},T} V_T T,S_{T-1}=s).\label{eq:lem1eq2}
\end{alignat}
\end{lemma}
Using Lemma~\ref{lemma1} we now replace the mutual information terms in \eqref{no1}~-~\eqref{no5} and
define the following variables for $j\in\{1,2\}$ and $s\in\mathcal{S}$.
\begin{align}
u^{(j)}_s&=I(U_{j,T};X_{T}|T,S_{T-1}=s)\\
z^{(j)}_s&=I(U_{j,T};X_T|U_{\bar{j},T}V_T T,S_{T-1}=s).
\end{align}
We have
\begin{align}
&R_j-\delta\leq\sum_{s\in\mathcal{S}}\pi_s (1-\epsj{s}) u^{(j)}_s, \quad j=1,2 \label{eq:conv_bound1}\\
&R_j-\delta\leq\sum_{s\in\mathcal{S}}\pi_s (1-\epsonetwo{s}) z^{(j)}_s,\quad  j=1,2. \label{eq:conv_bound2}
\end{align}

The following Lemma relates the parameters defined above and is proven in Appendix~\ref{appendix:lemma2}.
\begin{lemma}
\label{lemma2}
For every $j\in\{1,2\}$ and $s\in\mathcal{S}$, we have
\begin{align*}
&u^{(j)}_s+z^{(\bar{j})}_s\leq 1.
\end{align*}
\end{lemma}
Combining the above results and letting $\delta$ go to zero, $(R_1,R_2)$ can be achieved only if, for some variables $u_s^{(1)}$, $u_s^{(2)}$, $z_s^{(1)}$, $z_s^{(2)}$,  the following inequalities hold for all $j\in\{1,2\}$, $s\in\mathcal{S}$:
\begin{align}
&0\leq u^{(j)}_s, z^{(j)}_s\leq 1 \\
&u^{(j)}_s+z^{(\bar{j})}_s\leq 1 \label{tight}\\
&R_j\leq\sum_{s\in\mathcal{S}}\pi_s (1-\epsj{s}) u^{(j)}_s \\
&R_j\leq\sum_{s\in\mathcal{S}}\pi_s (1-\epsonetwo{s}) z^{(j)}_s .
\end{align}
The final step is to show that the above outer bound matches $ \bar{\mc C}_{\text{ fb}}^\text{mem}$ defined in \eqref{eq:posouter} - \eqref{eq:R2_constr2outer}.
This is done by noting that inequality \eqref{tight} can be made tight without changing the rate region. The equivalence of the two regions then becomes clear by setting 
$z^{(1)}_s=1-y_s$, $z^{(2)}_s=1-x_s$, $u^{(1)}_s=x_s$, and $u^{(2)}_s=y_s$.
%

%% file: schemes.tex
\section{Achievable Schemes}
\label{sec:ach}

\subsection{Queue and Flow Model}
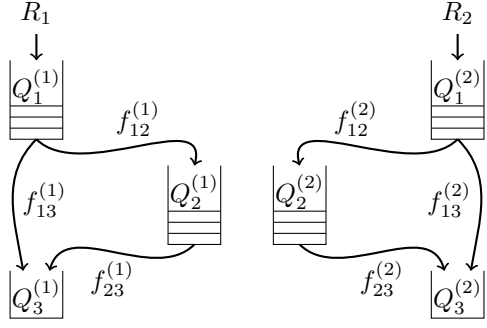
\begin{figure}[t]
\centering
\begin{tikzpicture}[scale=0.7]
\draw (0,0) -- ++(0,-1.5cm) -- ++(1.0cm,0) -- ++(0,1.5cm);
\foreach \i in {1,...,3}
  \draw (0,-1.5cm+\i*6pt) -- +(+1.0cm,0);

\draw (3,-2) -- ++(0,-1.5cm) -- ++(1.0cm,0) -- ++(0,1.5cm);
\foreach \i in {1,...,3}
  \draw (3,-3.5cm+\i*6pt) -- +(+1.0cm,0);

\draw (0,-4) -- ++(0,-0.9cm) -- ++(1.0cm,0) -- ++(0,0.9cm);

\draw (8,0) -- ++(0,-1.5cm) -- ++(1.0cm,0) -- ++(0,1.5cm);
\foreach \i in {1,...,3}
  \draw (8,-1.5cm+\i*6pt) -- +(+1.0cm,0);

\draw (5,-2) -- ++(0,-1.5cm) -- ++(1.0cm,0) -- ++(0,1.5cm);
\foreach \i in {1,...,3}
  \draw (5,-3.5cm+\i*6pt) -- +(+1.0cm,0);

\draw (8,-4) -- ++(0,-0.9cm) -- ++(1.0cm,0) -- ++(0,0.9cm);

\node at (0.5,-0.5) {$Q_1^{(1)}$}; 
\node at (8.5,-0.5) {$Q_1^{(2)}$}; 

\node at (0.5,-4.5) {$Q_3^{(1)}$};
\node at (8.5,-4.5) {$Q_3^{(2)}$};  
  
\node at (3.5,-2.5) {$Q_2^{(1)}$};
\node at (5.5,-2.5) {$Q_2^{(2)}$};


\draw[<-, thick] (0.5,0) -- +(0,0.5) node[above] {$R_1$};

\draw[->, thick] (0.5,-1.5) to [out=315,in=75] node[above] {$f_{12}^{(1)}$}(3.5,-2);
\draw[->, thick] (0.5,-1.5) to [out=225,in=100] node[right]{$f_{13}^{(1)}$} +(-0.25,-2.5);
\draw[->,thick] (3.5,-3.5) to [out=225,in=75] node[below]{$f_{23}^{(1)}$} +(-2.75,-0.5);

\draw[<-, thick] (8.5,0) -- +(0,0.5) node[above] {$R_2$};

\draw[->, thick] (8.5,-1.5) to [out=225,in=105] node[above] {$f_{12}^{(2)}$} (5.5,-2);
\draw[->, thick] (8.5,-1.5) to [out=315,in=80] node[left]{$f_{13}^{(2)}$} +(0.25,-2.5);
\draw[->,thick] (5.5,-3.5) to [out=315,in=105] node[below]{$f_{23}^{(2)}$} +(2.75,-0.5);

\end{tikzpicture}
\caption{Networked system of queues.}
\label{fig:queues}
\vspace*{-5mm}
\end{figure}

In this section, we develop codes that achieve the rate region $\underline{\mc C}_{\text{ fb}}^\text{mem}$. For this, we build on the idea of tracking packets that have been received at the wrong destination, as in \cite{georgiadis2009broadcast,6522177}. 
The transmitter has two buffers $Q_1^{(1)}$, $Q_1^{(2)}$ to store packets destined for $\text{Rx}_1$, $\text{Rx}_2$, respectively.
We consider dynamic arrivals, where packets for $\text{Rx}_1$, $\text{Rx}_2$ arrive in each slot according to a Bernoulli process with probability $R_1$, $R_2$, respectively. An analysis for more general arrival processes is possible. 
The transmitter has two additional buffers $Q_2^{(1)}$ (resp. $Q_2^{(2)}$) for packets that have already been sent, but have been received only by $\text{Rx}_2$ (resp. $\text{Rx}_1$). 
Hence buffer $Q_2^{(1)}$ contains packets that are destined for $\text{Rx}_1$ and have been received at $\text{Rx}_2$ but not at $\text{Rx}_1$, and vice versa for $Q_2^{(2)}$. These queues are empty before transmission begins.
Each user $j$, $j=1,2$, has a buffer $Q_3^{(j)}$ that collects desired packets. These buffers correspond to the system exit and are always empty.
The networked queuing system is shown in Fig.~\ref{fig:queues}.

Each receiver has an additional buffer (not depicted in Fig.~\ref{fig:queues}) that collects packets not intended for it, i.e. packets for the other user. Note that packets in this buffer are either also present in $Q_2^{(1)}$, $Q_2^{(2)}$, or have left the system. 

A packet for $\text{Rx}_j$ will only traverse buffers with superscript $j$, i.e. $Q_1^{(j)},Q_2^{(j)}$ or $Q_3^{(j)}$. 
In the following, slightly abusing the notation, we use $Q_{l,t}^{(j)}$ to denote the number of packets stored in buffer $Q_{l}^{(j)}$ in time slot $t$. Obviously, $Q_{l,t}^{(j)} \in \mc Q$ with $\mc Q = \{0,1,\ldots,\infty\}$. 
Define
\begin{align}
 \ve Q_t = \left( Q_{1,t}^{(1)}, Q_{2,t}^{(1)}, Q_{1,t}^{(2)}, Q_{2,t}^{(2)} \right) \in \mc Q^4.
\end{align}
Because $Q_3^{(1)}= Q_3^{(2)}=0$ by definition, the vector $\ve Q_t$ determines the queue state at time $t$.

If both $Q_2^{(1)}$ and $Q_2^{(2)}$ are nonempty, the transmitter can send the XOR combination of these packets.
If both users receive this coded packet, both can decode one desired packet and two packets per slot are delivered.
In general, the transmitter can select its action $A_t$ in slot $t$ from the set of actions $\mc A = \{1,2,3\}$ where
\begin{itemize}[leftmargin=+.5in]
 \item[$A_t=1$]corresponds to sending a packet for $\text{Rx}_1$ from $Q_1^{(1)}$, 
 \item[$A_t=2$]corresponds to sending a packet for $\text{Rx}_2$ from $Q_1^{(2)}$, 
 \item[$A_t=3$]corresponds to sending a coded packet.  
\end{itemize}
Actions at time $t$ are based on the \emph{current} queue state $\ve Q_t$ and the \emph{previous} channel state $S_{t-1}$.

Note that we permit actions from the action space $\mc A$ only. The corresponding stability region consists of all rate tuples $(R_1,R_2)$ for which all queues in the network are strongly stable \cite[Definition 3.1]{georgiadis2006resource}, i.e., if 
\begin{align}
 \limsup_{n\rightarrow \infty} \frac{1}{n} \sum_{t=1}^n \mathbb{E}[Q_t] < \infty. \label{eq:strong_stability}
\end{align}
A network is strongly stable if all queues are strongly stable \cite[Definition 3.2]{georgiadis2006resource}.
The algorithms developed in the following ensure network stability for rate pairs inside $\underline{\mc C}_{\text{ fb}}^\text{mem}$. The analysis is based on \cite{georgiadis2006resource, neely2010stochastic}, extended to incorporate the setup.

\subsection{Probabilistic Scheme}

Consider a strategy that bases decisions for actions only on the previous channel state $S_{t-1}$, but not on the queue state $\ve Q_t$.
These strategies are called S-only algorithms in \cite{neely2010stochastic}.
The decisions are random and independent from previous decisions, according to a probability distribution $P_{A_t|S_{t-1}}$ that does not depend on $t$.

Let $F_{lm,t}^{(j)}$ denote the number of packets that can travel from buffer $Q_l^{(j)}$ to $Q_m^{(j)}$ in slot $t$.
Clearly, $F_{lm,t}^{(j)}$ depends on the action chosen in slot $t$. Recall that $Z_{j,t}$ is equal to one if an erasure occurs at time $t$ for Rx$_j$ and is zero otherwise. So, we have
\begin{align}
 F_{12,t}^{(1)} = \truth\{A_t=1\} Z_{1,t}(1-Z_{2,t}).
 \label{eq:def_F12t1}
\end{align}
The long-term average rate $f_{12}^{(1)}$ is bounded by  
\begin{align}
f_{12}^{(1)} &\leq \lim_{n\rightarrow \infty} \frac{1}{n} \sum_{t=1}^n F_{12,t}^{(1)}=  \mathbb{E}[F_{12,t}^{(1)}], \label{eq:flow_bound_avg}
\end{align}
where the expectation in \eqref{eq:flow_bound_avg} is taken over the random previous channel state $S_{t-1}$, the random erasure events and the possibly random action $A_t$. Equality in \eqref{eq:flow_bound_avg} is achieved if $Q_{1,t}^{(1)}>0$ whenever $A_t=1$.
Similarly, we have
\begin{align}
 f_{13}^{(1)} &\leq \mathbb{E}[F_{13,t}^{(1)}],\quad F_{13,t}^{(1)} = \truth\{A_t=1\}(1-Z_{1,t})\\
 f_{23}^{(1)} &\leq \mathbb{E}[F_{23,t}^{(1)}],\quad F_{23,t}^{(1)} = \truth\{A_t=3\}(1-Z_{1,t})
 \label{eq:def_F23t1}
\end{align}
and correspondingly for the flows to $\text{Rx}_2$.

Thus, with this scheme, rate tuples $(R_1,R_2)$ can be achieved if there is a distribution $P_{A_t|S_{t-1}}$ such that $\forall~j \in \{1,2\}$:
\begin{align}
  R_j &\leq f_{13}^{(j)} + f_{12}^{(j)} \label{eq:rate_bound} \\
 f_{12}^{(j)} &\leq f_{23}^{(j)}  \label{eq:flow_cons} \\
  f_{12}^{(j)} &\leq \sum_{s\in \mc S} \pi_s  P_{A_t|S_{t-1}}(j|s) (\epsj{s} - \epsonetwo{s})   \label{eq:bound12} \\
  f_{13}^{(j)} &\leq \sum_{s\in \mc S} \pi_s P_{A_t|S_{t-1}}(j|s) (1-\epsj{s} )  \label{eq:bound13} \\
  f_{23}^{(j)} &\leq \sum_{s\in \mc S} \pi_s P_{A_t|S_{t-1}}(3|s) (1-\epsj{s} ).\label{eq:bound23}
\end{align}
Note that the region described by \eqref{eq:rate_bound} - \eqref{eq:bound23} is equivalent to the rate region $\underline{\mc C}_\text{fb}^\text{mem}$ described in \eqref{eq:R1_constr1} -  \eqref{eq:R2_constr2}. This may be seen by setting $P_{A_t|S_{t-1}}(1|s)=1-y_s$, $P_{A_t|S_{t-1}}(2|s)=1-x_s$, $P_{A_t|S_{t-1}}(3|s)=x_s+y_s-1$ and eliminating the flow variables $f_{lm}^{(j)}$. 
Whereas \eqref{eq:rate_bound} - \eqref{eq:bound23} is a \emph{maximum flow} formulation, \eqref{eq:R1_constr1} -  \eqref{eq:R2_constr2} describes the dual \emph{minimum cut} formulation.
Note that inequality \eqref{eq:pa3_pos} ensures that $P_{A_t|S_{t-1}}(3|s)\geq0$. This inequality is implicitly required in this approach but does not appear in the outer bound $\bar{\mc C}_\text{fb}^\text{mem}$.

\subsection{Deterministic Scheme}
\label{sec:ach_det}
In the probabilistic scheme,  
actions are chosen depending only on the channel state, so it can happen that there is no packet to transmit because the corresponding buffer is empty.
This can be avoided by a max-weight backpressure-like algorithm \cite{tassiulas1992stability,tassiulas1993dynamic,neely2005dynamic,georgiadis2006resource,neely2010stochastic} that bases its actions on both queue and channel states.

In each slot $t$,
the action maximizing the weight function in \eqref{eq:maxweight} is chosen: 
   \begin{align}
   A_t = \arg \max_{A\in \mc A} \sum_{j=1}^2 \truth\{A=j\} \left( [1-\epsj{s}] Q_1^{(j)} + (\epsj{s} - \epsonetwo{s})  (Q_1^{(j)}-Q_2^{(j)}) \right) + \truth\{A=3\} [1- \epsj{s}]  Q_2^{(j)} \label{eq:maxweight}
 \end{align}
Table~\ref{tab:det_algo} lists the weights for each action depending on the current queue state $\ve Q_t$ and the previous channel state $S_{t-1}=s$.

\begin{table}[t]
\centering
 \begin{tabular}{|c|c|l|}
 \hline
 Action $A_t$ & Weight depending on $\ve Q_t$ and $S_{t-1}=s$  \\ \hline
 $1$ &  $[1-\epsone{s}] Q_1^{(1)} + \epsonenottwo{s}  (Q_1^{(1)}-Q_2^{(1)}) $\\
 $2$ &  $[1-\epstwo{s}] Q_1^{(2)} + \epsnotonetwo{s}  (Q_1^{(2)}-Q_2^{(2)}) $\\
 $3$ &  $[1- \epsone{s}]  Q_2^{(1)} + [1- \epstwo{s}]  Q_2^{(2)}$\\\hline
 \end{tabular}
\caption{Deterministic scheme. }
\label{tab:det_algo}
\vspace*{-5mm}
\end{table}

\begin{prop}
The max-weight strategy in 
Table~\ref{tab:det_algo}
stabilizes all queues in the network for every rate pair $(R_1+\bar\delta, R_2+\bar\delta) \in \underline{\mc C}_{\text{ fb}}^\text{mem}$, $\bar \delta>0$.
\label{prop:maxweight}
\end{prop}
The proof is given in Appendix \ref{sec:proof_prop_maxweight}.

The rule in \eqref{eq:maxweight} ensures that actions are chosen only if the corresponding queues contain packets.

The proof uses a $T$-slot Lyapunov drift analysis similar to \cite{neely2010stochastic} but has to be adapted so that it takes into account only the previous channel states instead of the current one. This difference changes parts of the proof and the corresponding max-weight policy.
In the model of \cite{pantelidou2009cross}, the authors deal with correlated channels but have the \emph{current} channel state (or an estimate of it) available for the current decision.
Similarly, in \cite{neely2005dynamic, tassiulas1997scheduling}, the current channel state is available at the transmitter.
In \cite{li2011exploiting, li2013network} the authors focus on obtaining channel state information in a scenario that is related to the case of hidden states, however without permitting coding operations.
Similarly, \cite{ying2011throughput} investigates the case of delayed channel state information for general networks, without permitting coding operations.
During the preparation of this work we were informed that a similar approach was analyzed in \cite{Kuo_Wang2014} in a parallel line of work. More powerful coding actions are permitted in \cite{Kuo_Wang2014} that allow to close the gap to the outer bound.

%% file: results.tex
\section{Discussion}
\label{sec:comparison}

\begin{figure}[t]
 \centering
 \begin{tikzpicture}[->,>=stealth',shorten >=1pt,auto,node distance=40mm,
  thick,main node/.style={circle,fill=gray!20,draw},every node/.style={scale=1}]
%
  \node[main node] (GG) {$\GG$};
  \node[main node] (GB) [below of=GG] {$\GB$};
  \node[main node] (BG) [right of=GG] {$\BG$};
  \node[main node] (BB) [below of=BG] {$\BB$};

  \path[every node/.style={font=\sffamily\small}]
    (GG) edge [bend right=10] node [below,pos=0.7] {$b_1 b_2$} (BB)
        edge [bend right=20] node[above,rotate=90] {$(1-b_1)b_2$} (GB)
        edge [loop above] node {$(1-b_1)(1-b_2)$} (GG)
        edge [bend left=20] node {$b_1(1-b_2)$} (BG)
    (BG) edge [bend left=20] node [below,rotate=90] {$(1-g_1)b_2$} (BB)
         edge [bend left=10] node [below, pos=0.7] {$g_1 b_2$} (GB)
         edge [bend left=20] node [above] {$g_1 (1-b_2)$} (GG)
         edge [loop above] node {$(1-g_1)(1-b_2) $} (BG)
    
    (GB) edge [bend right=20] node [below] {$b_1(1-g_2)$} (BB)
         edge [bend left=10] node [left, pos=0.8] {$b_1 g_2$} (BG)
         edge [bend right=20] node [above,rotate=90] {$(1-b_1)g_2$} (GG)
         edge [loop below] node {$(1-b_1)(1-g_2)$} (GB)    
         
    (BB) edge [bend right=20] node {$ g_1(1-g_2)$} (GB)
         edge [bend right=10] node[right,pos=0.8] {$g_1 g_2 $} (GG)
         edge [bend left=20] node [below,rotate=90] {$(1-g_1)g_2 $} (BG)
         edge [loop below] node {$(1-g_1)(1-g_2)$} (BB);
\end{tikzpicture}
\caption{Markov Chain of channel state space $\mc S$ with transition probabilities.}
\label{fig:states}
\end{figure}
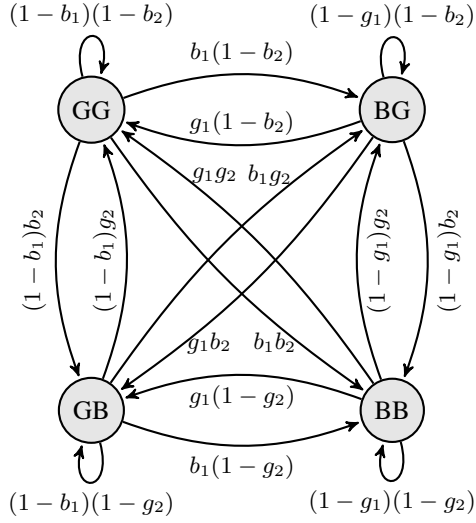

Consider the Gilbert-Elliot model of Example \ref{exampleGilbert}. We assume that the individual channels to users $1$ and $2$ are both Gilbert-Elliot channels with states $\G$ and $\B$. The broadcast channel state space is therefore given by $\mc S = \{\GG,\GB, \BG, \BB\}$ where $\G$ and $\B$ respectively refer to a good and bad state at each user. 
Transitions from  state $B$ to state $G$ occur with probability $g_j$ for user $j$, $j=1,2$. 
Similarly, a transition from state $G$ to state $B$ occurs with probability $b_j$ for user $j$.
For simplicity, these transitions are assumed to be independent across the two users.
The corresponding finite state Markov chain is summarized in Fig.~\ref{fig:states}.
The average (long-term) erasure probability at user user $j$ is given by
\begin{align}
 \epsilon_j = \frac{b_j}{g_j + b_j}\cdot
\end{align}

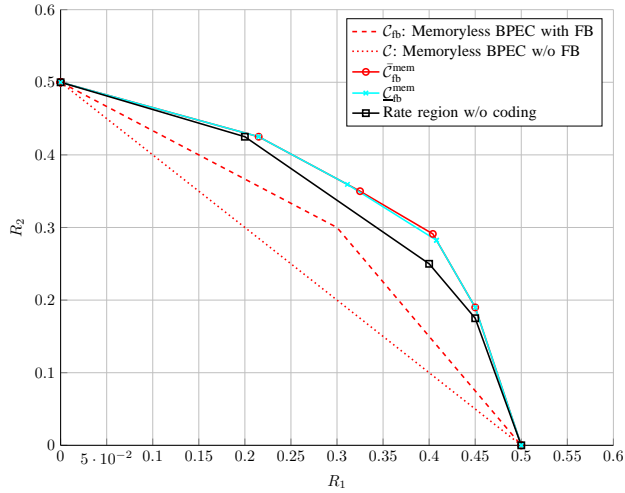
\begin{figure}[t]
 \centering
\input{./figs/RateRegions_e1=0.5,e2=0.5,g1=0.2,g2=0.3.tex}
\caption{Bounds on the Capacity region for $\epsilon_1=0.5$, $\epsilon_2=0.5$, $g_1=0.2$, $g_2=0.3$. In this case $\underline{\mc C}_{\text{fb}}^{\text{mem}}$ and $\bar{\mc C}_{\text{fb}}^{\text{mem}}$ almost match.
}
\label{fig:RateReg1}
\vspace*{-3mm}
\end{figure}

Fig.~\ref{fig:RateReg1} shows the capacity region for a channel with parameters $\epsilon_1=0.5$, $\epsilon_2=0.5$, $g_1=0.2$, $g_2=0.3$. In this figure we compare the bounds on the capacity region with that of a memoryless channel with the same average erasure probability (with and without feedback). We also show the rate region that is achieved by a simple scheme that does not permit coding across the two messages \cite{heindlmaier2014netcod}. This helps distinguishing the gains due to channel memory and the gains due to coding.   

\subsection{Combination of Memoryless Strategies}
Looking at the the characterization of $\underline{\mc C}_{\text{fb}}^{\text{mem}}$ in \eqref{eq:pos} - \eqref{eq:R2_constr2}, one may wonder if this rate-region can be attained simply by a combination of memoryless capacity achieving schemes. Let $\mc C_{\text{fb}}(s)$, $s\in \mc S$, denote the capacity region of a memoryless BPEC with feedback and erasure probabilities $\Pr[\ve Z_t|S_{t-1}=s]$. Capacity achieving algorithms  for memoryless BPEC with feedback are devised in \cite{georgiadis2009broadcast}. A combination of memoryless capacity achieving schemes may be described as follows: 
\begin{itemize}
 \item Choose fractions $\alpha_s\geq 0$ and $\beta_s\geq 0$ such that $\sum_{s\in \mc S} \alpha_s = \sum_{s\in \mc S} \beta_s=1$ and $(\alpha_s R_1, \beta_s R_2) \in \pi_s \mc C_{\text{fb}}(s)$, for all $s \in \mc S$. 
 \item Take $n \alpha_s R_1$ packets for $\text{Rx}_1$ and $n \beta_s R_2$ packets for $\text{Rx}_2$ to be transmitted only when the previous channel state is equal to $S_{t-1}=s$, $s \in \mc S$. For each previous state $s \in \mc S$, the transmitter chooses an optimal memoryless strategy (e.g., as devised in \cite{georgiadis2009broadcast}) corresponding to a memoryless BPEC channel with feedback and erasure probabilities $\Pr[\ve Z_t|S_{t-1}=s]$.
\end{itemize}

Using the above scheme, for large $n$, one can asymptotically achieve the performance of the memoryless strategy for each state $s$ with the corresponding capacity region $\mc C_{\text{fb}}(s)$. The overall rate region achievable by this strategy, called $\mc R_{\oplus}$, is thus a weighted combination of the individual memoryless rate regions (for each state $s$):
\begin{align}
 \mc R_{\oplus} = \bigoplus_{s \in \mc S} \pi_s \mc C_{\text{fb}}(s),
\end{align}
where $\oplus$ denotes the set addition operator\footnote{For example, $\pi_1 \mc R_1 \oplus \pi_2 \mc R_2 = \{\pi_1 \ve r_1 + \pi_2 \ve r_2 | \ve r_1 \in \mc R_1, r_2 \in \mc R_2 \} $.} (Minkowski sum).

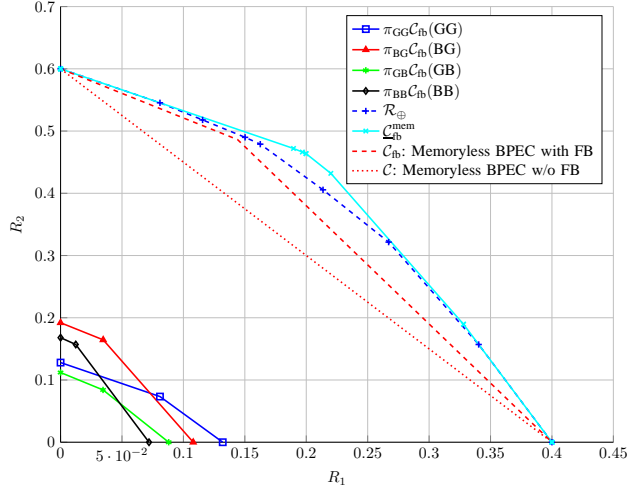
\begin{figure}[t]
 \centering
\input{./figs/Mink_sum_e1=0.6,g1=0.3,e2=0.4,g2=0.7.tex}
\caption{Individual rate regions and Minkowski sum $\mc R_{\oplus}$ for $\epsilon_1=0.6$, $\epsilon_2=0.4$, $g_1=0.3$, $g_2=0.7$. The region $\underline{\mc C}_{\text{fb}}^{\text{mem}}$ is strictly larger. For comparison, the corresponding capacity regions for memoryless channels with the same average erasure probability are shown for the cases with and without feedback. The difference between $\underline{\mc C}_{\text{fb}}^{\text{mem}}$ and $\bar{\mc C}_{\text{fb}}^{\text{mem}}$ is negligible in this case, so $\bar{\mc C}_{\text{fb}}^{\text{mem}}$ is omitted.
}
\label{fig:Minksum}
\vspace*{-3mm}
\end{figure}

We show in Fig.~\ref{fig:Minksum} that $\mc R_{\oplus}$ can be strictly smaller than $\underline{\mc C}_{\text{fb}}^{\text{mem}}$.%

\begin{remark}
Note that each memoryless rate region $\mc C_{\text{fb}}(s)$, $s\in \mc S$, is a polytope defined by linear inequalities.
However, the polytope generated by the Minkowski sum is \emph{not} equal to the one defined by the sum of the individual polytope constraints.
That would be the case, for example, if the memoryless rate regions were polymatroids, as pointed out in \cite[Chapter 15.3.3]{cover2006elements},\cite{traskov2012scheduling}. In that case, $\mc R_{\oplus}$ would be equal to $\underline{\mc C}_{\text{fb}}^{\text{mem}}$. However, this is not the case in general.
\end{remark}

\subsection{Delayed Feedback}
The result in Theorem~\ref{thm:capacity} extends to the scenario where feedback and channel state become available at the encoder with more than a single symbol-time delay. Consider a  delay of $d\geq1$ time units and call the achievable rate region $\underline{\mc C}_{\text{fb}}^{\text{mem}}(d)$. In the converse, one can obtain the corresponding bounds by replacing the sequences $S^{T-1}$, $Y_1^{T-1}$ $Z_1^{T-1}$, $Y_2^{T-1}$ and $Z_2^{T-1}$ with $S^{T-d}$, $Y_1^{T-d}$ $Z_1^{T-d}$, $Y_2^{T-d}$ and $Z_2^{T-d}$.

The bounds on the capacity region  $\underline{\mc C}_{\text{fb}}^{\text{mem}}(d)$ and $\bar{\mc C}_{\text{fb}}^{\text{mem}}(d)$ have thus a characterization as in \eqref{eq:posouter} - \eqref{eq:R2_constr2outer}, \eqref{eq:pos} - \eqref{eq:R2_constr2}, by redefining the erasure probabilities in \eqref{eq:def_epsilon} as
\begin{align}
 \epsonetwo{s}&=P_{\ve Z_t|S_{t-d}}(1,1|s),\quad
 &\epsonenottwo{s}&=P_{\ve Z_t|S_{t-d}}(1,0|s),\nonumber\\
 \epsnotonetwo{s}&=P_{\ve Z_t|S_{t-d}}(0,1|s),\quad  
  &\epsnotonenottwo{s} &=P_{\ve Z_t|S_{t-d}}(0,0|s),\nonumber\\
   \epsone{s}&=\epsonetwo{s}+\epsonenottwo{s},\quad 
  &\epstwo{s}&=\epsonetwo{s}+\epsnotonetwo{s}. \nonumber 
\end{align}
The corresponding deterministic achievable scheme as in Section~\ref{sec:ach_det} uses these redefined conditional erasure probabilities to obtain the same description as in Table~\ref{tab:det_algo}.

\begin{figure}[t]
\centering
\input{./figs/delay_regions_e1=0.6,e2=0.5,g1=0.1,g2=0.1.tex}
\caption{Capacity regions with delayed feedback, for $\epsilon_1=0.6$, $g_1=0.1$, $\epsilon_2=0.5$, $g_2=0.1$. The genie-aided outer bound corresponds to the case when all erasures are known ahead of time.}
\label{fig:delay}
\vspace*{-3mm}
\end{figure}
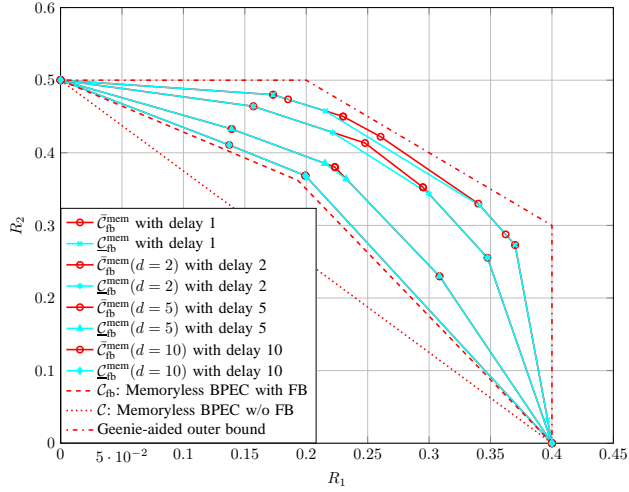

Fig.~\ref{fig:delay} shows the effect of feedback delay for a Gilbert-Elliot channel with parameters  $\epsilon_1=0.6$, $g_1=0.1$, $\epsilon_2=0.5$, $g_2=0.1$. One observes that delayed feedback shrinks both the outer and inner bounds, as the state information becomes less useful. After a feedback delay of $d=10$ time units for this example, the region $\underline{\mc C}_{\text{fb}}^{\text{mem}}(d=10)$ is almost the same as for the memoryless case. In general this depends on the convergence speed of the state Markov chain towards its stationary distribution.
It is interesting to see that, as $d$ increases the difference between $\underline{\mc C}_{\text{fb}}^{\text{mem}}$ and $\bar{\mc C}_{\text{fb}}^{\text{mem}}$ becomes smaller.
$\underline{\mc C}_{\text{fb}}^{\text{mem}}$ and $\bar{\mc C}_{\text{fb}}^{\text{mem}}$ match for the memoryless BPEC.

%% file: figs/RateRegions_e1=0.5,e2=0.5,g1=0.2,g2=0.3.tex
%
%
%
\definecolor{mycolor1}{rgb}{0,1,1}%
\begin{tikzpicture}[scale=0.6]

\begin{axis}[%
width=4.82222222222222in,
height=3.80333333333333in,
scale only axis,
xmin=0,
xmax=0.6,
xlabel={$R_1$},
xmajorgrids,
ymin=0,
ymax=0.6,
ylabel={$R_2$},
ymajorgrids,
axis x line*=bottom,
axis y line*=left,
legend style={draw=black,fill=white,legend cell align=left}
]
\addplot [
color=red,
dashed,
line width=1.0pt
]
table[row sep=crcr]{
0 0.5\\
0.3 0.3\\
0.5 0\\
};
\addlegendentry{$\mc C_{\text{fb}}$: Memoryless BPEC with FB};

\addplot [
color=red,
dotted,
line width=1.0pt
]
table[row sep=crcr]{
0 0.5\\
0.5 0\\
};
\addlegendentry{$\mc C$: Memoryless BPEC w/o FB};

\addplot [color=red,solid,line width=1.0pt,mark=o,mark options={solid}]
  table[row sep=crcr]{%
-6.3881585785075e-10	0.499999998130779\\
-2.13710755689323e-10	0.499999999398456\\
1.70510008890012e-10	0.499999999336781\\
2.21881249484746e-10	0.499999999328496\\
2.75490883577811e-10	0.499999999319605\\
3.31491344790491e-10	0.499999999310049\\
3.90131420224815e-10	0.499999999299615\\
4.51741009183104e-10	0.499999999288115\\
5.8451112006086e-10	0.499999999262883\\
6.59816385556455e-10	0.499999999245871\\
7.46618900385188e-10	0.499999999221409\\
1.58461724557996e-08	0.499999994231098\\
0.215000000131436	0.424999999812868\\
0.325000010300229	0.349999991654879\\
0.404020468021695	0.291154969886671\\
0.449999999967801	0.189999999846005\\
0.499999994022514	-2.564584586362e-09\\
0.499999999802543	4.94053364191727e-10\\
};
\addlegendentry{$\bar{\mc C}_{\text{fb}}^{\text{mem}}$};

\addplot [
color=mycolor1,
solid,
line width=1.0pt,
mark=x,
mark options={solid}
]
table[row sep=crcr]{
1.26301822268788e-10 0.499999999894753\\
0.214999997206266 0.425000000464778\\
0.215000002017283 0.424999998333691\\
0.311315164865004 0.359330568917867\\
0.408010680059939 0.282376502247954\\
0.408010682812185 0.282376496593052\\
0.450000000104885 0.189999999091287\\
0.499999999566823 1.33745843283371e-09\\
0.499999999831986 2.51077649096155e-10\\
};
\addlegendentry{$\underline{\mc C}_{\text{fb}}^{\text{mem}}$ };

\addplot [color=black,solid,line width=1.0pt,mark=square,mark options={solid}]
  table[row sep=crcr]{%
0	0.5\\
0.2	0.425\\
0.4	0.25\\
0.45	0.175\\
0.5	0\\
};
\addlegendentry{Rate region w/o coding};

\end{axis}
\end{tikzpicture}%

%% file: figs/Mink_sum_e1=0.6,g1=0.3,e2=0.4,g2=0.7.tex
%
%
%
\definecolor{mycolor1}{rgb}{0,1,1}%
\begin{tikzpicture}[scale=0.6]

\begin{axis}[%
width=4.82222222222222in,
height=3.80333333333333in,
scale only axis,
xmin=0,
xmax=0.45,
xlabel={$R_1$},
xmajorgrids,
ymin=0,
ymax=0.7,
ylabel={$R_2$},
ymajorgrids,
axis x line*=bottom,
axis y line*=left,
legend style={draw=black,fill=white,legend cell align=left}
]

\addplot [
color=blue,
solid,
line width=1.0pt,
mark=square,
mark options={solid}
]
table[row sep=crcr]{
0 0.128\\
0.0809186737881689 0.0733713594679029\\
0.132 0\\
};
\addlegendentry{$\pi{}_{\text{GG}}\mc{ C}_{\text{fb}}(\text{GG})$};

\addplot [
color=red,
solid,
line width=1.0pt,
mark=triangle*,
mark options={solid}
]
table[row sep=crcr]{
0 0.192\\
0.0347020148462354 0.164513255567338\\
0.108 0\\
};
\addlegendentry{$\pi{}_{\text{BG}}\mc{ C}_{\text{fb}}(\text{BG})$};

\addplot [
color=green,
solid,
line width=1.0pt,
mark=asterisk,
mark options={solid}
]
table[row sep=crcr]{
0 0.112\\
0.0345785670039232 0.0840173446211026\\
0.088 0\\
};
\addlegendentry{$\pi{}_{\text{GB}}\mc{C}_{\text{fb}}(\text{GB})$};

\addplot [
color=black,
solid,
line width=1.0pt,
mark=diamond,
mark options={solid}
]
table[row sep=crcr]{
0 0.168\\
0.0123622313450857 0.157046124124608\\
0.072 0\\
};
\addlegendentry{$\pi{}_{\text{BB}}\mc{C}_{\text{fb}}(\text{BB})$};

\addplot [
color=blue,
dashed,
line width=1.0pt,
mark=+,
mark options={solid}
]
table[row sep=crcr]{
0.4 0\\
0.340362231345086 0.157046124124608\\
0.267064246191321 0.321559379691946\\
0.213642813195244 0.405576724313048\\
0.162561486983413 0.478948083780951\\
0.150199255638328 0.489901959656344\\
0.115620688634404 0.517884615035241\\
0.0809186737881689 0.545371359467903\\
0 0.6\\
};
\addlegendentry{$\mc{R}_{\oplus}$};

\addplot [
color=mycolor1,
solid,
line width=1.0pt,
mark=x,
mark options={solid}
]
  table[row sep=crcr]{%
-2.20554111952931e-09	0.599999994315768\\
-2.17065155606877e-09	0.599999994577054\\
-2.12795018064704e-09	0.599999994820831\\
-2.07779714787859e-09	0.599999995048889\\
-1.47167629818701e-09	0.59999999775458\\
-1.723679358967e-11	0.599999999925199\\
7.68199913403489e-12	0.599999999928623\\
0.189600000728751	0.47199999930733\\
0.196832802429257	0.466271047146064\\
0.199830683135434	0.463720834318547\\
0.199830684957968	0.463720831498871\\
0.219999999898067	0.431999999883207\\
0.328000000151881	0.189599999312797\\
0.32800000328372	0.189599991150972\\
0.399999996247596	2.77485687205647e-10\\
0.399999999939633	-1.46718748261776e-12\\
0.399999999957988	1.54624368864376e-11\\
0.399999999958552	1.61638480378201e-11\\
0.399999999960768	1.9636958725755e-11\\
};
\addlegendentry{$\underline{\mc C}_{\text{fb}}^{\text{mem}}$ };

\addplot [
color=red,
dashed,
line width=1.0pt
]
table[row sep=crcr]{
0 0.6\\
0.144075829383886 0.486255924170616\\
0.4 0\\
};
\addlegendentry{$\mc C_{\text{fb}}$: Memoryless BPEC with FB };

\addplot [
color=red,
dotted,
line width=1.0pt
]
table[row sep=crcr]{
0 0.6\\
0.4 0\\
};
\addlegendentry{$\mc C$: Memoryless BPEC w/o FB  };


\end{axis}
\end{tikzpicture}%

%% file: figs/delay_regions_e1=0.6,e2=0.5,g1=0.1,g2=0.1.tex
%
%
%
\definecolor{mycolor1}{rgb}{0,1,1}%
\definecolor{mycolor2}{rgb}{1.00000,1.00000,0.00000}%

\begin{tikzpicture}[scale=0.6]

\begin{axis}[%
width=4.82222222222222in,
height=3.80333333333333in,
scale only axis,
xmin=0,
xmax=0.45,
xlabel={$R_1$},
xmajorgrids,
ymin=0,
ymax=0.6,
ylabel={$R_2$},
ymajorgrids,
legend style={at={(0.0,0.0)},anchor=south west,draw=black,fill=white,legend cell align=left}
]

\addplot [color=red,solid,line width=1.0pt,mark=o,mark options={solid}]
  table[row sep=crcr]{%
3.04262984945902e-09	0.499999999396795\\
3.39388148834496e-09	0.499999999551646\\
0.172999999833664	0.479999999964128\\
0.172999999840085	0.479999999961635\\
0.185170529989744	0.473594457811832\\
0.229999999691038	0.449999999413554\\
0.230000008201643	0.449999992117119\\
0.260477521688112	0.422152517327349\\
0.339999999938638	0.329999999827842\\
0.36230476488635	0.287620946489435\\
0.369999999951003	0.272999999880099\\
0.369999999970275	0.272999999727871\\
0.39999999907244	8.00312011627058e-09\\
0.399999999288013	4.51659832798046e-09\\
};
\addlegendentry{$\bar{\mc C}_{\text{fb}}^{\text{mem}}$ with delay 1};

\addplot [
color=mycolor1,
solid,
line width=1.0pt,
mark=x,
mark options={solid}
]
  table[row sep=crcr]{%
5.42115626001038e-09	0.499999999304043\\
0.172999999900408	0.479999999923169\\
0.172999999974989	0.479999999887034\\
0.215561608619569	0.457599153271377\\
0.215561610481258	0.457599152199471\\
0.341503094666189	0.327144119908171\\
0.369999999966637	0.272999999864625\\
0.370000000027495	0.272999999444877\\
0.396483251465987	0.0320024104856066\\
0.399999997397321	1.65580349119343e-08\\
};
\addlegendentry{$\underline{\mc C}_{\text{fb}}^{\text{mem}}$ with delay 1};


\addplot [color=red,solid,line width=1.0pt,mark=o,mark options={solid}]
  table[row sep=crcr]{%
4.08231694976058e-11	0.499999999930361\\
5.06030946312297e-11	0.499999999768872\\
4.85336916489221e-10	0.499999995856295\\
6.38944857092594e-10	0.499999995644111\\
0.156949999708745	0.463999999968596\\
0.156949999918903	0.463999999912291\\
0.24783566705575	0.413429922303068\\
0.294999999129591	0.352500000691768\\
0.295000000538491	0.352499998172797\\
0.347500000007089	0.25544999967901\\
0.399999994026922	7.9731871452382e-10\\
0.399999999806744	6.44442089969832e-10\\
};
\addlegendentry{$\bar{\mc C}_{\text{fb}}^{\text{mem}}(d=2)$ with delay 2};

\addplot [
color=mycolor1,
solid,
line width=1.0pt,
mark=asterisk,
mark options={solid}
]
table[row sep=crcr]{
8.88397894701065e-10 0.499999999498033\\
9.72639632900947e-10 0.499999999711467\\
0.156949996778929 0.464000000604486\\
0.15695000070875 0.463999999347534\\
0.221431505722674 0.428121571690184\\
0.299718334955756 0.343777820578036\\
0.347499999349731 0.25545000091145\\
0.347499999771156 0.255449999135915\\
0.347500002477089 0.255449986920886\\
0.399999999878136 2.30874239548158e-10\\
};
\addlegendentry{$\underline{\mc C}_{\text{fb}}^{\text{mem}}(d=2)$ with delay 2};

%
%
%
%

\addplot [color=red,solid,line width=1.0pt,mark=o,mark options={solid}]
  table[row sep=crcr]{%
-1.50906845700072e-09	0.499999997511849\\
-5.46040636317802e-10	0.499999999062048\\
4.65750413930488e-10	0.499999999697495\\
0.139243078518205	0.432767999438278\\
0.139243089807093	0.432767992681422\\
0.22330308055164	0.380311162785594\\
0.223303096693208	0.380311152696028\\
0.223303098752961	0.380311150290188\\
0.308476559627929	0.229918522813686\\
0.30847656309711	0.229918515046333\\
0.399999998953125	-2.09943472329055e-10\\
0.399999999851802	4.24361934481254e-12\\
0.399999999926315	7.17781563353004e-11\\
0.39999999992684	6.44289482876204e-11\\
};
\addlegendentry{$\bar{\mc C}_{\text{fb}}^{\text{mem}}(d=5)$ with delay 5};

\addplot [
color=mycolor1,
solid,
line width=1.0pt,
mark=triangle*,
mark options={solid}
]
table[row sep=crcr]{
1.50950512569393e-10 0.499999999859184\\
0.139243083699421 0.432767997283583\\
0.215242765842038 0.385341125219978\\
0.215242766900777 0.385341124175141\\
0.215242767836298 0.385341123230204\\
0.232548136867466 0.363986986464548\\
0.308476562467742 0.229918518641014\\
0.399999998808526 1.24366086078451e-09\\
0.399999999882481 1.72509471699247e-10\\
};
\addlegendentry{$\underline{\mc C}_{\text{fb}}^{\text{mem}}(d=5)$ with delay 5};

\addplot [color=red,solid,line width=1.0pt,mark=o,mark options={solid}]
  table[row sep=crcr]{%
-8.54477464479197e-10	0.49999999857806\\
-4.87467403248454e-10	0.499999999505802\\
2.99853839674791e-10	0.499999999731377\\
0.137299156056543	0.410737418615813\\
0.199269570285195	0.368531509390085\\
0.199269574063453	0.36853150659245\\
0.199269574714597	0.368531505866853\\
0.199269575388053	0.368531504755233\\
0.199269849586659	0.368531023962609\\
0.399999994552454	1.02901706044989e-08\\
0.399999999005104	-2.6770734240511e-10\\
0.399999999879181	-2.23184387637687e-11\\
0.399999999943554	5.55554213743648e-11\\
};
\addlegendentry{$\bar{\mc C}_{\text{fb}}^{\text{mem}}(d=10)$ with delay 10};

\addplot [
color=mycolor1,
solid,
line width=1.0pt,
mark=diamond*,
mark options={solid}
]
table[row sep=crcr]{
8.97265272955639e-11 0.499999999911609\\
2.04855639298161e-09 0.499999997285639\\
0.137299085813 0.4107374608347\\
0.19914318855396 0.368617583144707\\
0.200209704805009 0.366899387297235\\
0.399999971905306 5.34943439534591e-08\\
0.399999999012641 9.56804233676889e-10\\
0.399999999874006 1.71880204557762e-10\\
};
\addlegendentry{$\underline{\mc C}_{\text{fb}}^{\text{mem}}(d=10)$ with delay 10};

\addplot [
color=red,
dashed,
line width=1.0pt
]
table[row sep=crcr]{
0 0.5\\
0.193103448275862 0.362068965517241\\
0.4 0\\
};
\addlegendentry{$\mc C_{\text{fb}}$: Memoryless BPEC with FB};

\addplot [
color=red,
dotted,
line width=1.0pt
]
table[row sep=crcr]{
0 0.5\\
0.4 0\\
};
\addlegendentry{$\mc C$: Memoryless BPEC w/o FB};

\addplot [
color=red,
dash pattern=on 1pt off 3pt on 3pt off 3pt,
line width=1.0pt
]
table[row sep=crcr]{
0 0.5\\
0.2 0.5\\
0.4 0.3\\
0.4 0\\
};
\addlegendentry{Geenie-aided outer bound};

\end{axis}
\end{tikzpicture}%

%% file: appendix.tex
\section{Proof of Lemma \ref{lemma1}}
\label{appendix:lemma1}
In the following, we prove \eqref{eq:lem1eq1} of Lemma \ref{lemma1}. \eqref{eq:lem1eq2} 
is derived similarly.
\begin{align}
I&(U_{j,T};Y_{j,T}|T,S_{T-1}=s)=I(U_{j,T};Y_{j,T}Z_{j,T}|T,S_{T-1}=s)\nonumber\\
&=I(U_{j,T};Z_{j,T}|T,S_{T-1}=s)+I(U_{j,T};Y_{j,T}|Z_{j,T}T,S_{T-1}=s)\nonumber\\
&\stackrel{(a)}{=}I(U_{j,T};Y_{j,T}|Z_{j,T}T,S_{T-1}=s)\nonumber\\
&= \epsj{s} I(U_{j,T};Y_{j,T}=E|T,S_{T-1}=s,Z_{j,T}=1)+\left(1-\epsj{s}\right)I(U_{j,T};X_{T}|T,S_{T-1}=s,Z_{j,T}=0)\nonumber\\
&\stackrel{(b)}{=}\left(1-\epsj{s}\right)I(U_{j,T};X_{T}|T,S_{T-1}=s)\label{eq5}.
\end{align}
In the above chain of equalities, $(a)$ and $(b)$ both follow by the Markov chain $Z_{j,T}-S_{T-1}T-U_{1,T}U_{2,T}V_{T}X_T$ and because $I(U_{j,T};Y_{j,T}=E|T,S_{T-1}=s,Z_{j,T}=1)=0$.

\section{Proof of Lemma \ref{lemma2}}
\label{appendix:lemma2}
First note that $I(U_{1,T} U_{2,T} V_T; X_T|T, S_{T-1}=s) \leq 1$ because $H(X_T|T,S_{T-1}=s)\leq 1$. Applying the chain rule, we obtain 
\begin{align}
1\geq &I(U_{1,T} U_{2,T} V_T; X_T|T, S_{T-1}=s)  \nonumber \\
 =&I(U_{1,T}; X_T|T,S_{T-1}=s) + I( V_T; X_T|U_{1,T} T, S_{T-1}=s) + I(U_{2,T}; X_T|U_{1,T} V_T T ,S_{T-1}=s)\nonumber \\
 \geq &u_s^{(1)} + I( V_T; X_T|U_{1,T} T, S_{T-1}=s) + z_s^{(2)} \nonumber \\
 \geq &u_s^{(1)} + z_s^{(2)}.
\end{align}
Similarly, $u_s^{(2)} + z_s^{(1)}\leq 1$.

\section{Proof of Proposition \ref{prop:maxweight}}
\label{sec:proof_prop_maxweight}
The objective is to  show that the max-weight criterion in \eqref{eq:maxweight} strongly stabilizes all queues in the network, as defined in \eqref{eq:strong_stability}. An approach using Lyapunov-drift theory will lead to this result.

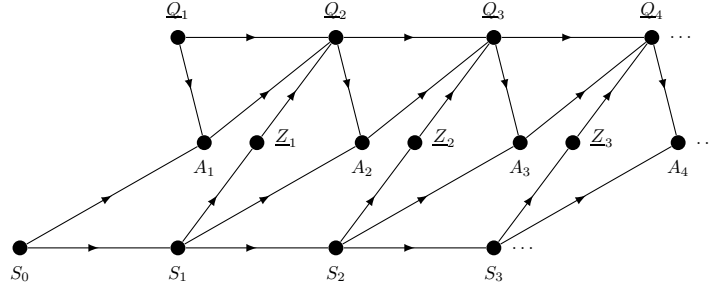
\begin{figure}[ht]
\centering
  \tikzset{>=latex}
\begin{tikzpicture}[scale = 0.7, every node/.style={scale=0.7}]
\input{./figs/fdg_queues.tex}
 \end{tikzpicture}
\caption{Bayesian network of the queuing system. 
Actions at time $t$ are permitted to depend on the previous channel state $S_{t-1}$ and the current buffer state $\ve Q_t$. Note that $\ve Q_t$ denotes the buffer state \emph{before} action $A_t$ is executed.}
\label{fig:fdg_queues}
\end{figure}

Let $F_{01,t}^{(j)}$ denote the indicator random variable if a packet arrived in queue $Q_1^{(j)}$ during time slot $t$. $F_{01,t}^{(j)}$ is independent from all other random variables in the system, so $\mathbb E[F_{01,t}^{(j)}]=R_j$. 
Recall the definitions of $F_{lm,t}^{(j)}$ in \eqref{eq:def_F12t1} - \eqref{eq:def_F23t1} and let $\tilde F_{lm,t}^{(j)} = F_{lm,t}^{(j)} \truth\{Q_l^{(j)}>0\}$. $\tilde F_{lm,t}^{(j)}$ describes the actual number of packets traveling from queues $Q_{l}^{(j)}$ to $Q_{m}^{(j)}$ during time slot $t$.
 The dynamics of queues $Q_{1,t}^{(j)}$, $Q_{2,t}^{(j)}$ are given by
 \begin{align}
  Q_{1,t+1}^{(j)} &= Q_{1,t}^{(j)} - \tilde F_{12,t}^{(j)} - \tilde F_{13,t}^{(j)} + F_{01,t}^{(j)} \nonumber \\
  &= [Q_{1,t}^{(j)} - F_{12,t}^{(j)} - F_{13,t}^{(j)}]^+ + F_{01,t}^{(j)}\\
  Q_{2,t+1}^{(j)} &= Q_{2,t}^{(j)} - \tilde F_{23,t}^{(j)} + \tilde F_{12,t}^{(j)} \nonumber \\
    &\leq [Q_{2,t}^{(j)} - F_{23,t}^{(j)}]^+ + F_{12,t}^{(j)} \label{eq:queue_update2}
 \end{align}
where $[x]^+ = \max(x,0)$. There is an inequality rather than an equality in \eqref{eq:queue_update2} because $Q_{1,t}^{(j)}$ could be empty and thus $\tilde F_{12,t}^{(j)} \leq F_{12,t}^{(j)}$ (i.e., no packet could travel from buffer $Q_1^{(j)}$ to $Q_2^{(j)}$ although $F_{12,t}^{(j)}=1$).
The flow variables $F_{lm,t}^{(j)}$ depend on the action $A_t$ and the erasures $\ve Z_t$, so the queue state $\ve Q_{t+1}$ is a function of $\ve Q_t$, $A_t$ and $\ve Z_t$. Actions are restricted to depend only on the current queue state and on the previous channel state. This is a possibly suboptimal choice, but we will see that this choice is sufficient to achieve rate points in $\underline{\mc C}_{\text{ fb}}^\text{mem}$. All dependencies are depicted in the Bayesian network in Fig.~\ref{fig:fdg_queues}.

Define the Lyapunov function $L(\ve Q_t)$ as
\begin{align}
 L(\ve Q_t) = \sum_{j=1}^2 \sum_{l=1}^2 \left( Q_{l,t}^{(j)} \right)^2
\end{align}
and the $T$-slot conditional Lyapunov drift $\Delta(\ve Q_t)$ as
\begin{align}
 \Delta(\ve Q_t) = \mathbb{E}\left[ L(\ve Q_{t+T}) - L(\ve Q_{t}) \left|\ve Q_t \right. \right], \label{eq:condLyap}
\end{align}
where the expectation is with respect to the possibly random actions $A_t,A_{t+1},\ldots,A_{t+T-1}$, erasures $\ve Z_t, \ve Z_{t+1},\ldots,\ve Z_{t+T-1}$, previous states $S_{t-1}, S_{t},\ldots,S_{t+T-2}$ and queues $\ve Q_{t+1},\ve Q_{t+2},\ldots,\ve Q_{t+T}$.
$\Delta(\ve Q_t)$ is a measure of the expected reduction or increase of the queue lengths from slot $t$ to slot $t+T$, conditioned on $\ve Q_t$, and will be useful to prove strong stability.

Split $\Delta(\ve Q_t)$ into the following telescoping sum
\begin{align}
  \Delta(\ve Q_t) =& \left.\mathbb{E}\left[\sum_{\tau=t}^{t+T-1} L(\ve Q_{\tau+1}) - L(\ve Q_\tau) \right\vert \ve Q_t  \right] \nonumber\\
  =&  \sum_{\tau=t}^{t+T-1} \mathbb{E}\left[ L(\ve Q_{\tau+1}) - L(\ve Q_\tau) \left| \ve Q_t \right. \right].
  \label{eq:telescopingsum}
  \end{align}
  Note that the individual expectation terms of the sum in \eqref{eq:telescopingsum} depend on the conditioning $\ve Q_t$ only through $\ve Q_\tau$ and $S_{\tau-1}$, as $\ve Q_{\tau+1} A_\tau \ve Z_\tau - \ve Q_\tau S_{\tau-1} - \ve Q_t$ forms a Markov chain for $\tau\geq t$.
  Hence, one can use the law of total expectation and write 
\begin{align}
 &\Delta(\ve Q_t)= \label{eq:total_exp}
 \sum_{\tau=t}^{t+T-1} \mathbb E \Big[ \mathbb E \left[ L(\ve Q_{\tau+1}) - L(\ve Q_\tau) \big| \ve Q_\tau S_{\tau-1}  \right] \Big| \ve Q_t \Big].
\end{align}

We will bound the individual terms inside the inner expectation of \eqref{eq:total_exp}.
To this end, we can use \cite[Lemma 4.3]{georgiadis2006resource}, which states that
for any nonnegative numbers $v,u,\mu,\alpha$ satisfying $v \leq [u-\mu]^+ + \alpha$, we have
\begin{align}
v^2 &\leq u^2 + \mu^2 + \alpha^2 - 2u(\mu - \alpha).
\end{align}
We apply this lemma and combine it with the fact that $(F_{lm,\tau}^{(j)})^2 = F_{lm,\tau}^{(j)} \leq 1$ because $F_{lm,\tau}^{(j)}$ is either $1$ or $0$ and obtain the following bound: 
%
 \begin{align}
 L(\ve Q_{\tau+1})-L(\ve Q_\tau)&\!\leq
 \!\  \sum_{j=1}^2  F_{01,\tau}^{(j)} \!+\! 2 F_{12,\tau}^{(j)} \!+\! F_{13,\tau}^{(j)} \!+\!F_{23,\tau}^{(j)} \!-\! 2 Q_{1,\tau}^{(j)} \left(F_{12,\tau}^{(j)}\!+\! F_{13,\tau}^{(j)} \!-\!F_{01,\tau}^{(j)} \right) \!-\! 2 Q_{2,\tau}^{(j)} \left( F_{23,\tau}^{(j)} \!-\! F_{12,\tau}^{(j)}  \right) \nonumber \\
 &\leq 
 10 - 2\sum_{j=1}^2   Q_{1,\tau}^{(j)} \left( F_{12,\tau}^{(j)}+ F_{13,\tau}^{(j)} -F_{01,\tau}^{(j)} \right) +  Q_{2,\tau}^{(j)} \left( F_{23,\tau}^{(j)} - F_{12,\tau}^{(j)}  \right). \label{eq:flow_bound}
\end{align}

The bound in \eqref{eq:flow_bound} is used to develop a bound on $\Delta(\ve Q_t)$.
First, we insert \eqref{eq:flow_bound} back into \eqref{eq:total_exp} to obtain 
 \begin{align}
 \Delta(\ve Q_t)  &\stackrel{}{\leq} \sum_{\tau=t}^{t+T-1}\mathbb E\left[\mathbb E\left[10 - 2\sum_{j=1}^2   Q_{1,\tau}^{(j)} \left( F_{12,\tau}^{(j)}+ F_{13,\tau}^{(j)} -F_{01,\tau}^{(j)} \right) +  Q_{2,\tau}^{(j)} \left( F_{23,\tau}^{(j)} - F_{12,\tau}^{(j)}  \right)\Bigg| \ve Q_\tau S_{\tau-1}  \right] \Bigg| \ve Q_t \right]\\
  &\stackrel{(a)}{\leq} \sum_{\tau=t}^{t+T-1} 10 - \mathbb E \left[ 2\sum_{j=1}^2   Q_{1,\tau}^{(j)} \Big( \Pr[A_\tau=j|\ve Q_\tau S_{\tau-1}] \left( 1- \epsonetwo{S_{\tau-1}}  \right) - R_j \Big)  \right. 
 \nonumber\\
 &\qquad\Bigg. +  Q_{2,\tau}^{(j)} \Big( \Pr[A_\tau=3|\ve Q_\tau S_{\tau-1}] (1-\epsj{S_{\tau-1}}) - \Pr[A_\tau=j|\ve Q_\tau S_{\tau-1}] (\epsj{S_{\tau-1}} - \epsonetwo{S_{\tau-1}})  \Big) \Bigg|\ve Q_t \Bigg] \label{eq:drift_eq}.
\end{align}
In the above chain of inequalities, step ($a$) is derived as follows.
For $j\in \{1,2\}$, we have 
\begin{align}
&\mathbb E[F_{12,\tau}^{(j)}+F_{13,\tau}^{(j)}|\ve Q_{\tau} S_{\tau-1}]  \nonumber \\
=& \mathbb E[ \truth\{A_\tau=j\} \left( Z_{j,\tau} (1-Z_{\bar{j},\tau}) + 1-Z_{j,\tau}\right) |\ve Q_{\tau} S_{\tau-1} ] \nonumber \\
=&\Pr[A_\tau=j|\ve Q_\tau,S_{\tau-1} ] \cdot  \left(1-\Pr[\ve Z_{\tau}=(1,1)|S_{\tau-1} ]\right) \nonumber\\
=&\Pr[A_\tau=j|\ve Q_\tau,S_{\tau-1} ] \cdot  \left( 1- \epsonetwo{S_{\tau-1}} \right),
\end{align}
because $\ve Z_\tau - S_{\tau-1} - A_\tau \ve Q_\tau$ forms a Markov chain. 
Corresponding steps apply for the other expressions inside the expectation, leading to \eqref{eq:drift_eq}.

Note that the criterion in \eqref{eq:maxweight} finds the tightest upper bound on $\Delta(\ve Q_t)$ in \eqref{eq:drift_eq}:
The distribution $P_{A_\tau|\ve Q_\tau S_{\tau-1}}$ that maximizes the expression inside the conditional expectation for every outcome of $\ve Q_\tau$ and $S_{\tau-1}$ also minimizes the upper bound in \eqref{eq:drift_eq}.
The associated optimization problem is a linear program, only constrained by conditions that $P_{A_\tau|\ve Q_\tau S_{\tau-1}}$ must be a probability distribution. The optimizer of a linear program lies at the boundary of the constraint set, and thus the optimal conditional distribution is  deterministic, leading to the max-weight criterion in \eqref{eq:maxweight}.

\begin{remark}
 Note that
 \begin{itemize}
  \item the criterion in \eqref{eq:maxweight} does not depend on $\tau$,
  \item actions are chosen only if the corresponding queues are nonempty, so no transmissions are wasted and $\tilde{F}_{lm,\tau}^{(j)} = F_{lm,\tau}^{(j)}$, unless all queues are empty.
 \end{itemize}
\end{remark}
The criterion in \eqref{eq:maxweight} finds the tightest upper bound in \eqref{eq:drift_eq} under the assumption that actions can depend on $Q_\tau$ and $S_{\tau-1}$. Hence, any stationary probabilistic scheme that bases its decisions only on $S_{\tau-1}$, according to a distribution $P_{A_\tau|S_{\tau-1}}$, leads to a looser upper bound on $\Delta(\ve Q_t)$ than the one ensured by the criterion in \eqref{eq:maxweight}. This is stated in \eqref{eq:drift_eq_proba} and serves as starting point to further bound $ \Delta(\ve Q_t)$ as follows. The individual  steps are explained below.
\allowdisplaybreaks

 \begin{align}
 \Delta(\ve Q_t) 
 &\stackrel{}{\leq} \sum_{\tau=t}^{t+T-1} 10 - \mathbb E  \left[ 2\sum_{j=1}^2   Q_{1,\tau}^{(j)} \Big( \Pr[A_\tau=j|S_{\tau-1}] \left( 1- \epsonetwo{S_{\tau-1}}  \right) - R_j \Big)  \right. 
 \nonumber\\
 &\qquad\Bigg. +  Q_{2,\tau}^{(j)} \Big( \Pr[A_\tau=3| S_{\tau-1}] (1-\epsj{S_{\tau-1}}) - \Pr[A_\tau=j| S_{\tau-1}] (\epsj{S_{\tau-1}} - \epsonetwo{S_{\tau-1}})  \Big) \Bigg|\ve Q_t \Bigg] \label{eq:drift_eq_proba}\\
 &\stackrel{(b)}{=} \sum_{\tau=t}^{t+T-1} 10 - \mathbb E \left[ 2\sum_{j=1}^2   Q_{1,\tau}^{(j)} \flowdiv{Q_1}{(j)}{S_{\tau-1}}+  Q_{2,\tau}^{(j)} \flowdiv{Q_2}{(j)}{S_{\tau-1}} \Bigg|\ve Q_t \right] \label{eq:drift_eq_simpl}\\
 &\stackrel{(c)}{\leq} \sum_{\tau=t}^{t+T-1} 10 + 8 (\tau-t) - \mathbb E \left[ 2\sum_{j=1}^2   Q_{1,t}^{(j)} \flowdiv{Q_1}{(j)}{S_{\tau-1}}+  Q_{2,t}^{(j)} \flowdiv{Q_2}{(j)}{S_{\tau-1}} \Bigg|\ve Q_t \right] \label{eq:drift_eq_with_Qbound}\\
 &\stackrel{(d)}{=} 10T + 4T(T-1) - 2 \sum_{j=1}^2 \sum_{\tau=t}^{t+T-1} \sum_{s\in \mc S} \Pr[S_{\tau-1}=s|\ve Q_t]  \left(   Q_{1,t}^{(j)} \flowdiv{Q_1}{(j)}{s}+  Q_{2,t}^{(j)} \flowdiv{Q_2}{(j)}{s} \right) \\
 &\stackrel{(e)}{=}10T + 4T(T-1)- 2 \sum_{j=1}^2  \sum_{s\in \mc S}   \left(  Q_{1,t}^{(j)} \flowdiv{Q_1}{(j)}{s}+  Q_{2,t}^{(j)} \flowdiv{Q_2}{(j)}{s} \right) \sum_{\tau=t}^{t+T-1} \Pr[S_{\tau-1}=s|\ve Q_t]\\
 &\stackrel{(f)}{\leq} 10T+4T^2 - 2T\sum_{j=1}^2 Q_{1,t}^{(j)} \sum_{s\in \mc S} \left( \pi(s) \flowdiv{Q_1}{(j)}{s} - \varepsilon    \right) + Q_{2,t}^{(j)}  \sum_{s\in \mc S}\left( \pi(s) \flowdiv{Q_2}{(j)}{s} - \varepsilon   \right)\\
 &\stackrel{(g)}{\leq}10T + 4T^2 -2T(\delta-\varepsilon |\mc S|) \sum_{j=1}^2 \left( Q_{1,t}^{(j)} + Q_{2,t}^{(j)}\right).\label{eq:final_bound_Delta}
\end{align}

To simplify notation in step $(b)$, we write the flow divergence of buffer $Q_l^{(j)}$ as $\flowdiv{Q_l}{(j)}{s}$.
The flow divergence \cite[Chapter 1.1.2]{bertsekas1998network} is the average number of packets that can depart from buffer $Q_l^{(j)}$ minus the average number of packets that can arrive at $Q_l^{(j)}$, given that the previous channel state is $s$ and the probabilistic scheme according to some $P_{A_\tau|S_{\tau-1}}$ is used.
Hence,
\begin{align}
 \flowdiv{Q_1}{(j)}{s} &= \mathbb E\left[  F_{12,\tau}^{(j)}+ F_{13,\tau}^{(j)} -F_{01,\tau}^{(j)} | S_{\tau-1}=s\right] \\
 \flowdiv{Q_2}{(j)}{s} &= \mathbb E\left[  F_{23,\tau}^{(j)} -F_{12,\tau}^{(j)} | S_{\tau-1}=s\right].
\end{align}
Note that the flow divergence $\flowdiv{Q_l}{(j)}{s}$ is an average and does not depend on the buffer level $Q_{l,\tau}^{(j)}$.

For step $(c)$ we follow similar steps as in \cite[Sect. 4.9]{neely2010stochastic}: The buffer level $Q_{l,\tau}^{(j)}$ can decrease by at most one per slot, because at most one packet can depart in each time slot:
\begin{align}
 \qquad Q_{l,\tau}^{(j)} &\geq Q_{l,t}^{(j)} - (\tau -t), \quad \text{for }\tau \geq t \label{eq:lb_Ql}.
\end{align}
One obtains \eqref{eq:drift_eq_with_Qbound}, where the expression inside the expectation does not depend on $\ve Q_\tau$ anymore.
Steps $(d)$ and $(e)$ write out the expectation and rearrange terms.

The constant $T$ can be chosen large enough such that 
\begin{align}
 \left| \pi(s) - 1/T \sum_{\tau=t}^{t+T-1} \Pr[S_{\tau-1}=s|\ve Q_t] \right| \leq \varepsilon, \quad \forall s \in \mc S \label{eq:converge_to_steady}
\end{align}
for some $\varepsilon >0$. 
Such a value of $T$ always exists if the Markov chain of the channel state process is irreducible and aperiodic\footnote{Aperiodicity is not necessarily required due to the Ces\`{a}ro mean in \eqref{eq:converge_to_steady}, but this is beyond the scope of this work.}: 
In this case the steady-state distribution $\pi$ is unique (see, e.g., \cite[Theorem 4.3.1]{gallager2013stochastic}) and the distribution $P_{S_{\tau-1}|\ve Q_t}$ converges to $\pi$ for any initial distribution $P_{S_{t-1}|\ve Q_t}$, $\tau > t$. If $P_{S_{\tau-1}|\ve Q_t}$ converges to $\pi$, so does the Ces\`{a}ro mean in \eqref{eq:converge_to_steady}.
The constant $T$ is thus related to the mixing time of the channel state Markov chain.

We replace $\sum_{\tau=t}^{t+T-1} \Pr[S_{\tau-1}=s|\ve Q_t]$ by its lower bound $T(\pi(s)-\varepsilon)$ for step $(f)$, 
where we have also used the fact that $\flowdiv{Q_l}{(j)}{s} \leq 1$ for all $l$, $j$ and $s$.

For step $(g)$, note that if the rate pair is in the interior of the rate region defined by \eqref{eq:rate_bound} - \eqref{eq:bound23}, i.e. if $(R_1+\bar{\delta}, R_2+\bar{\delta}) \in \underline{\mc C}_{\text{ fb}}^\text{mem}$, then there exists a constant $\delta >0$ that goes to zero when $\bar{\delta} \rightarrow 0$ such that
\begin{align}
\sum_{s\in \mc S} \pi(s) \flowdiv{Q_l}{(j)}{s}\geq \delta, \quad \forall~l\in \{1,2\},~j\in \{1,2\},
\end{align}
where $\delta$ should be chosen such that $\delta > \varepsilon |\mc S|$.

Using the result in \eqref{eq:final_bound_Delta} and the law of total expectation, we can bound
\begin{align}
 \mathbb E \left[ L(\ve Q_{t+T}) - L(\ve Q_t) \right] = &\mathbb E \bigl[ \Delta(\ve Q_t) \big| \ve Q_t \bigr] \nonumber \\
  \leq  14T^2 - 2T(\delta-\varepsilon |\mc S|) & \sum_{j=1}^2 \mathbb E\left[ Q_{1,t}^{(j)} + Q_{2,t}^{(j)} \right].
\end{align}
Summing over all time slots $t=1,\ldots,n$ yields
\begin{align}
\sum_{t=1}^{n} &\mathbb E\left[ L(\ve Q_{t+T})- L(\ve Q_t) \right] = \sum_{t=1}^T \mathbb E\left[L(\ve Q_{t+n}) - L(\ve Q_t)  \right]  \nonumber \\\leq &14 n T^2 - 2 T (\delta-\varepsilon |\mc S|)   \sum_{t=1}^{n} \sum_{j=1}^2 \mathbb E\left[Q_{1,t}^{(j)} + Q_{2,t}^{(j)} \right].
 \end{align}
 Rearranging terms gives
 \begin{align}
 \frac{1}{n} \sum_{t=1}^n \sum_{j=1}^2 \mathbb E[Q_{1,t}^{(j)} + Q_{2,t}^{(j)}] &\leq \frac{7T}{\delta - \varepsilon |\mc S|} + \frac{\sum_{t=1}^T \mathbb E[L(\ve Q_{t}) ]}{2(\delta - \varepsilon|\mc S|)T n},
\end{align}
and taking a $\limsup$ on both sides proves strong stability of the queuing network, given that $1/T\sum_{t=1}^T \mathbb E[L(\ve Q_{t}) ]<\infty$. This is true if the constant $T$ is finite and $\mathbb E[L(\ve Q_1)]<\infty$.
This proves Proposition~\ref{prop:maxweight}.

%% file: figs/fdg_queues.tex
\node at (2,-5) [fill,circle, inner sep = 1mm] (Q1) {};
\node [above=0.5mm of Q1] () {$\ve Q_{1}$};

\node at (5,-5) [fill,circle, inner sep = 1mm] (Q2) {};
\node [above=0.5mm of Q2] () {$\ve Q_{2}$};

\node at (8,-5) [fill,circle, inner sep = 1mm] (Q3) {};
\node [above=0.5mm of Q3] () {$\ve Q_{3}$};

\node at (11,-5) [fill,circle, inner sep = 1mm] (Q4) {};
\node [above=0.5mm of Q4] () {$\ve Q_{4}$};
\node [right=0.5mm of Q4] () {$\ldots$};

\node at (2.5,-7) [fill,circle, inner sep = 1mm] (A1) {};
\node [below=0.5mm of A1] () {$A_{1}$};

\node at (5.5,-7) [fill,circle, inner sep = 1mm] (A2) {};
\node [below=0.5mm of A2] () {$A_{2}$};

\node at (8.5,-7) [fill,circle, inner sep = 1mm] (A3) {};
\node [below=0.5mm of A3] () {$A_{3}$};

\node at (11.5,-7) [fill,circle, inner sep = 1mm] (A4) {};
\node [below=0.5mm of A4] () {$A_{4}$};
\node [right=0.5mm of A4] () {$\ldots$};

\node at (3.5,-7) [fill,circle, inner sep = 1mm] (Z1) {};
\node [right=0.5mm of Z1] () {$\ve Z_{1}$};

\node at (6.5,-7) [fill,circle, inner sep = 1mm] (Z2) {};
\node [right=0.5mm of Z2] () {$\ve Z_{2}$};

\node at (9.5,-7) [fill,circle, inner sep = 1mm] (Z3) {};
\node [right=0.5mm of Z3] () {$\ve Z_{3}$};

\node at (-1,-9) [fill,circle, inner sep = 1mm] (S0) {};
\node [below=0.5mm of S0] () {$S_{0}$};

\node at (2,-9) [fill,circle, inner sep = 1mm] (S1) {};
\node [below=0.5mm of S1] () {$S_{1}$};

\node at (5,-9) [fill,circle, inner sep = 1mm] (S2) {};
\node [below=0.5mm of S2] () {$S_{2}$};

\node at (8,-9) [fill,circle, inner sep = 1mm] (S3) {};
\node [below=0.5mm of S3] () {$S_{3}$};
\node [right=0.5mm of S3] () {$\ldots$};

\draw[->-=.5] (S0) -- (S1);
\draw[->-=.5] (S1) -- (S2);
\draw[->-=.5] (S2) -- (S3);

\draw[->-=.5] (S0) -- (A1);
\draw[->-=.5] (S1) -- (A2);
\draw[->-=.5] (S2) -- (A3);
\draw[->-=.5] (S3) -- (A4);

\draw[->-=.5] (S1) -- (Z1);
\draw[->-=.5] (S2) -- (Z2);
\draw[->-=.5] (S3) -- (Z3);

\draw[->-=.5] (Q1) -- (Q2);
\draw[->-=.5] (Q2) -- (Q3);
\draw[->-=.5] (Q3) -- (Q4);

\draw[->-=.5] (Q1) -- (A1);
\draw[->-=.5] (Q2) -- (A2);
\draw[->-=.5] (Q3) -- (A3);
\draw[->-=.5] (Q4) -- (A4);

\draw[->-=.5] (A1) -- (Q2);
\draw[->-=.5] (A2) -- (Q3);
\draw[->-=.5] (A3) -- (Q4);

\draw[->-=.5] (Z1) -- (Q2);
\draw[->-=.5] (Z2) -- (Q3);
\draw[->-=.5] (Z3) -- (Q4);